\documentclass[9.5pt,journal,compsoc]{IEEEtran}
\ifCLASSINFOpdf
\else
\fi
\usepackage{amsmath}
\usepackage{graphicx}
\usepackage{booktabs}
\usepackage{multirow}
\usepackage{flowchart}
\usepackage{times}
\usepackage{url}  
\usepackage{mathrsfs}
\usepackage{enumitem}
\usepackage[caption=false,font=normalsize,labelfont=sf,textfont=sf]{subfig}
\usepackage{url}
\begin{document}
\setlength{\baselineskip}{11.5pt}
\title{Retrieving Similar Trajectories from Cellular Data at City Scale}
% a structure at a ~ so this ke-eps-converted-to.pdf an author's name from being broken across
%% note need leading \protect in front of \\ to get a newline within \thanks as

\author{Zhihao Shen,
		Wan Du,~\IEEEmembership{Member,~IEEE,}
		Xi Zhao,~\IEEEmembership{Member,~IEEE,}
		and~Jianhua Zou~\IEEEmembership{Member,~IEEE,} % <-this % stops a space

\IEEEcompsocitemizethanks{\IEEEcompsocthanksitem Z. Shen and J. Zou is with the School of Electronic and Information Engineering, Xi'an Jiaotong University, and Shaanxi Engineering Research Center of Medical and Health Big Data, Xi'an 710049, China.\protect\\
E-mail: szh1095738849@stu.xjtu.edu.cn, jhzou@sei.xjtu.edu.cn
\IEEEcompsocthanksitem W. Du is with the Department of Computer Science and Engineering, University of California, Merced, CA 95340 USA.\protect\\
E-mail: wdu3@ucmerced.edu% <-this % stops an unwanted space
\IEEEcompsocthanksitem X. Zhao is with the School of Management, Xi'an Jiaotong University, and the Key Lab of the Ministry of Education for process control \& Efficiency Egineering, Xi'an 710049, China.\protect\\
E-mail: Zhaoxi1@mail.xjtu.edu.cn
%\IEEEcompsocthanksitem J. Zou is with the School of Electronic and Information Engineering, Xi'an Jiaotong University, Xi'an 710049, China.\protect\\
%E-mail: jhzou@sei.xjtu.edu.cn
}

\thanks{
% Manuscript received April 19, 2019. 
%revised August 26, 2015.
Wan Du is the first corresponding author and Xi Zhao is the second corresponding author.}}


\markboth{IEEE TRANSACTIONS ON MOBILE COMPUTING}
%,~Vol.~?, No.~?, ?~2019}%
{Shell \MakeLowercase{\textit{et al.}}: Bare Demo of IEEEtran.cls for IEEE Journals}




\IEEEtitleabstractindextext{%
\begin{abstract}
Retrieving similar trajectories from a large trajectory dataset is important for a variety of applications, like transportation planning and mobility analysis. Unlike previous works based on fine-grained GPS trajectories, this paper investigates the feasibility of identifying similar trajectories from cellular data observed by mobile infrastructure, which provide more comprehensive coverage. To handle the large localization errors and low sample rates of cellular data, we develop a holistic system, \textit{cellSim}, which seamlessly integrates map matching and similar trajectory search. A set of map matching techniques are proposed to transform cell tower sequences into moving trajectories on a road map by considering the unique features of cellular data, like the dynamic density of cell towers and bidirectional roads. To further improve the accuracy of similarity search, map matching outputs $M$ trajectory candidates of different confidence, and a new similarity measure scheme is developed to process the map matching results. Meanwhile, $M$ is dynamically adapted to maintain a low false positive rate of the similarity search, and two pruning schemes are proposed to minimize the computation overhead.
Extensive experiments on a large-scale cellular dataset and real-world trajectories of 1701 km reveal that cellSim provides high accuracy (precision 62.4\% and recall of 89.8\%).
\end{abstract}

\begin{IEEEkeywords}
Trajectory similarity, Map matching, Cellular data, Human mobility
\end{IEEEkeywords}}

\maketitle

\IEEEdisplaynontitleabstractindextext

% \fi
\IEEEpeerreviewmaketitle

\IEEEraisesectionheading{\section{Introduction}}
%GPS-enabled IoT (Internet of Things) devices and location-enabled mobile applications are widely adopted,
\IEEEPARstart{R}{etrieving} similar trajectories is to search in a large dataset for the trajectories that have a similar movement pattern. 
This is essential for many trajectory processing tasks (e.g., clustering~\cite{lee2007trajectory,chen2018online}, classification~\cite{lee2008traclass}) and many applications (e.g., human mobility analysis~\cite{luxiaoni,naserian2018framework,thuillier2018clustering,cao2019habit2vec} and transportation planning~\cite{zhang2016multicalib,meng2017city}). 
Many works have been proposed to define new similarity measures~\cite{dtw,lcss} or implement search operations in parallel~\cite{vldb}.
However, most existing works are designed for fine-grained GPS trajectories~\cite{yuan2010t,naboulsi2016large}, which can only cover a small population or space in a city, as GPS sensors are not installed in all vehicles or enabled by all mobile users.

This paper investigates the feasibility of finding a set of trajectories within a large-scale dataset that are close to a given query trajectory in time and space domain.
Benefiting from the pervasive usage of mobile phones and the extensive coverage of cellular networks, cellular data can provide a more comprehensive coverage in terms of population and space.
% Since many mobile phone apps, like YouTube or Google map, keep accessing the cellular infrastructure, the network packets are transferred through the associated cell towers frequently.
% Such routing can be recorded by mobile carriers.
Many anonymized cellular datasets \cite{zhang2014exploring,Feng2017A} have been released by mobile carriers. 
They are composed of time-series sequences, each of which is a series of time-stamped cell tower locations.   
They are processed for a variety of applications \cite{naboulsi2016large}, e.g., human mobility \cite{nature2009} and mobile network analysis~\cite{Li2017A}. 
% However, few works have been conducted for trajectory similarity search. 
% Beside the conventional applications of trajectory similarity search, our results can also be used by mobile carriers for network analysis, i.e., why do two co-moving persons associate with two different cell towers?
However, existing trajectory similarity measures cannot be simply applied due to the large localization error and low sample rate of cellular data, compared with GPS data. For example, two persons may associate with different cell towers that may be hundreds of meters apart when moving together; and the issue can become worse if their phones subscribe to different mobile carriers. 
%It is hard, if not impossible, to find an appropriate similarity threshold to classify the distance of two cell tower sequences.

We develop a holistic system, named as~\textit{cellSim}, which effectively searches for similar trajectories, which are close to the query trajectory in the space-time domain, in a large-scale cellular dataset of multiple carriers. 
First, we apply a Hidden Markov Model (HMM) to map-match each cell tower sequence in the cellular dataset to the trajectory. Then, we further output multiple trajectory candidates as the results of map matching to handle the low sample rate problem in the cellular data. Finally, a novel trajectory similarity measure is proposed to find similar trajectories. In this way, the trajectories having similarities larger than similarity threshold will be classified as the similar trajectories. 

We first map-match cell tower sequences into trajectories on a road map to avoid the location ambiguity of cellular data of multiple carriers.
Most existing map matching methods are designed to process GPS trajectories~\cite{Wei2013Map,Wei2012Fast} or cellular beacon data~\cite{thiagarajan2011accurate,paek2011energy}. Accurate localization can be acquired with the cellular beacon data, since multiple beacons from different cell towers can be received by a mobile phone simultaneously at a time. 
However, the cellular data used in this work can only provide coarse location information, as we only know the associated cell tower at every location. 
To implement more accurate map matching, a set of techniques are developed on top of a HMM-based map matching method while considering the unique features of our cellular data. The proposed algorithm can automatically adapt to the variety of cell tower density and the local properties of roads, including topological structure, road type and speed limit. 

We further explore reducing the low sample rate problem in the cellular data by seamlessly incorporating map matching into the trajectory similarity search. 
In particular, our map matching outputs top $M$ trajectory candidates, which probably include the true trajectory or one trajectory close to it. To compare two cell tower sequences, our similarity search performs pairwise comparisons between their $M$ trajectories while considering each candidate’s confidence.
\textit{If two persons are traveling together, their true trajectories must have the highest similarity}. We choose the highest similarity from the $M^2$ similarity results as the final result.
With $M$ trajectory candidates for each cell tower sequence, we increase the probability of finding truly similar trajectories, but we also increase the probability of finding false similar trajectories.
We propose an adaption algorithm of $M$ to minimize the false positive rate. 
In addition, two pruning techniques are developed to reduce the computation overhead of the similarity search.

\begin{figure}[tbp]
	\centerline{\includegraphics[width=3in,height=1in]{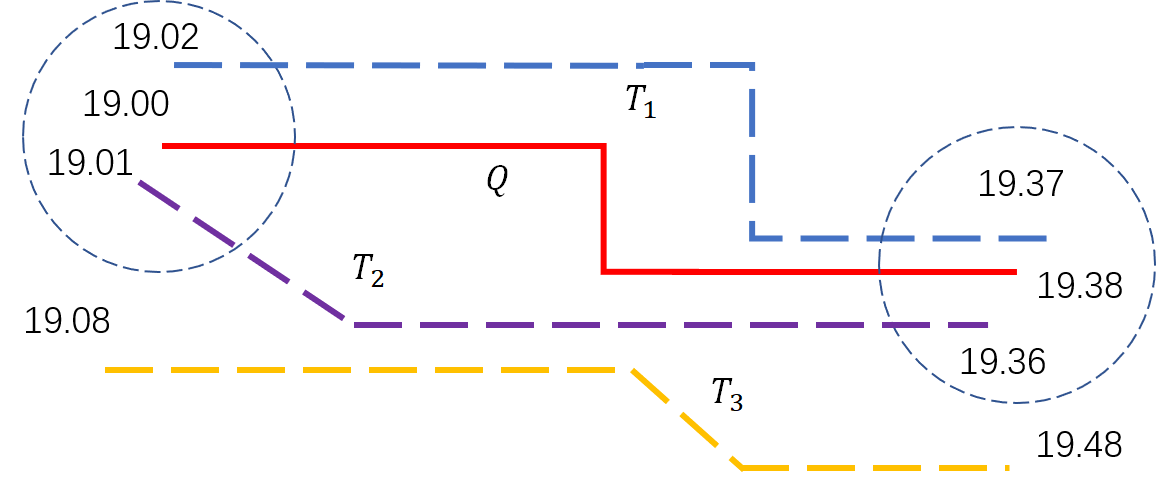}}	
	\caption{An example of similar trajectory search.}
	\vspace{-0.2in}
	\label{figSimilarityExample}
\end{figure}

Although the above map matching method could mitigate the location ambiguity of cellular data, its results still suffer from location deviations caused by the cellular data. 
We further design a trajectory similarity measure to better  measure the similarity between two trajectories generated from our cellular data. 
Most existing works~\cite{dtw,erp} measure the distance of two trajectories (e.g., Euclidean distance or edit distance). 
They work well for GPS data with low localization errors. 
However, due to the variety of cell tower density, some locations have large localization errors within our generated trajectories, which make the results of distance-based trajectory similarity measures vary drastically. It is hard to find a similarity threshold to identify similar trajectories.
On the contrary, our trajectory similarity measure first uses a sliding time window to extract road segments of two trajectories within the same time interval. Then, we calculate the similarities by comparing the spatially-overlapping ratio in all sliding windows.
By doing so, it reduces the impact of location deviations in trajectories, which increases its ability to tolerate noise. 
%
%By doing so, our method considers both the time information of two trajectories and their spatial distance on a road map in a continuous manner.  
%By doing so, the location deviations of the trajectories will have limited impact on the accuracy of searching similar trajectories. 

To the best of our knowledge, we are the first to develop a  system that extracts similar trajectories from a large-scale cellular dataset.
We implement cellSim in Hadoop~\cite{hadoop} with 3 master nodes and 10 slave nodes.  
We collect real-world trajectories of 1701 km by volunteers that move together as ground truth. 
For each trajectory in our ground truth data, we search for the other similar trajectories in the cellular dataset. 
Experiment results reveal that cellSim provides a precision and recall of 62.4\% and 89.8\% respectively, corresponding to a performance gain of 88.5\% and 65.7\% over the state-of-the-art solution. 
% In particular, our proposed map matching improves the conventional methods by 22.3\% and 6.7\%, and our new trajectory similarity search outperforms the existing schemes by 22.1\% and 8.1\%.
With our pruning scheme, cellSim only uses 51 seconds to search in the large-scale cellular dataset for one query. 

% In summary, this paper makes the following contributions.
% \begin{itemize}[leftmargin=0.2in]
%  	\item To the best of our knowledge, we are the first to develop a holistic system that extracts similar moving trajectories from large-scale cellular datasets. 
%  	\item To handle the location ambiguity of cellular data in retrieving moving trajectories, we develop innovative designs in both map matching and similarity searching, as well as the combination of both technologies.
%  	\item We implement cellSim in Hadoop and experimentally demonstrate its effectiveness  and efficiency.
% \end{itemize}

%to the journal version, we may give some possible reason to explain: why do two adjacent users associate with two different cell towers? 

\vspace{-0.1in}
\section{Motivation}
In this section, we introduce the trajectory similarity
search problem and our cellular dataset. 

\begin{figure}[tbp]
	\centering
	\subfloat[Cell towers.]{
		\label{figDensityCellTower} %% label for first subfigure
		\includegraphics[width=1.5in,height=1in]{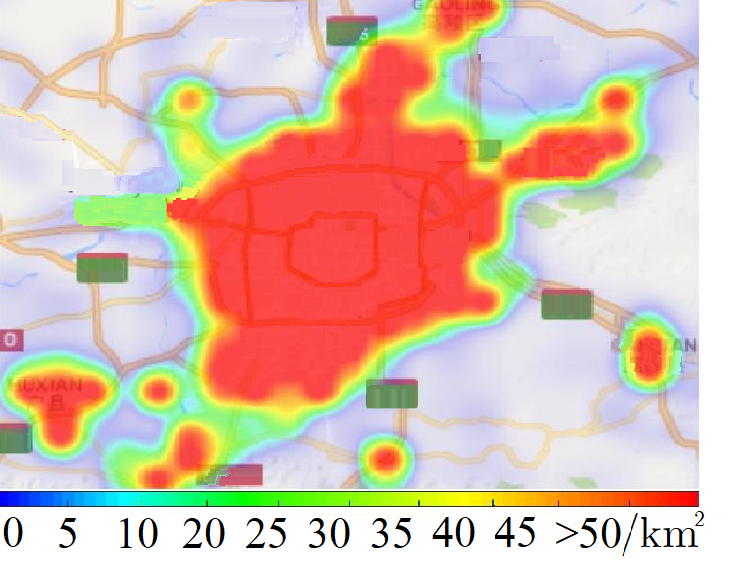}}
	\subfloat[Mobile users.]{
		\label{figDensityUsers} %% label for second subfigure
		\includegraphics[width=1.5in,height=1in]{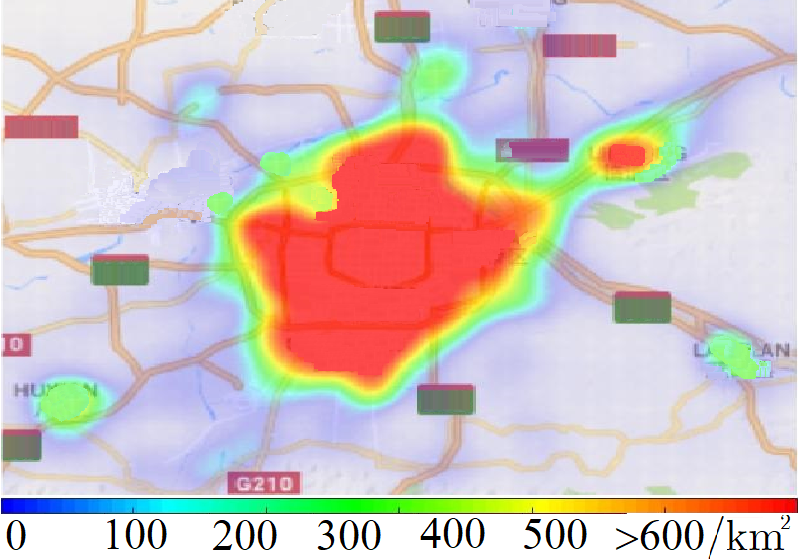}}
	\vspace{-0.1in}
	\caption{The density of cell towers and mobile users.}
	\vspace{-0.2in}
	\label{fig2Xian}
\end{figure}

% We also provide some preliminary processing results to demonstrate
% the challenges in using cellular data for extracting similar trajectories.
\vspace{-0.1in}
\subsection{Trajectory similarity search}
\label{subsecTrajectorySimilaritySearch}
% They cannot directly work on the raw cellular trajectories in general.
%why so many metrics have been proposed?
A trajectory is a sequence of locations (cell tower location or GPS point) to which an object travels, i.e., $T=p_1, p_2, ..., p_n$, where $p_i$ is a point $(x_i, y_i, t_i)$ with the location of $(x_i, y_i)$ at the time $t_i$.
In this paper, we define trajectory similarity search as the problem of finding a set of trajectories within a large-scale dataset that are close to a given query trajectory in the space-time domain.
%the similarities are larger than the similarity threshold. 
Existing trajectory similarity search methods  \cite{Chen2009Design,xie2017distributed,dtw} are designed for fine-grained GPS trajectories with low localization error and high sample rate. 
They normally have two steps, i.e., first extracting the trajectories that have the similar start and end time with the query trajectory, and then aligning the sample points using their timestamps to compare the spatial distance between each pairs of sample points.

\begin{figure}[tbp]
	\centering
	\subfloat[Single carrier]{
		\label{figCellTowerLocationsSame} %% label for first subfigure
		\includegraphics[width=1.5in,height=1.3in]{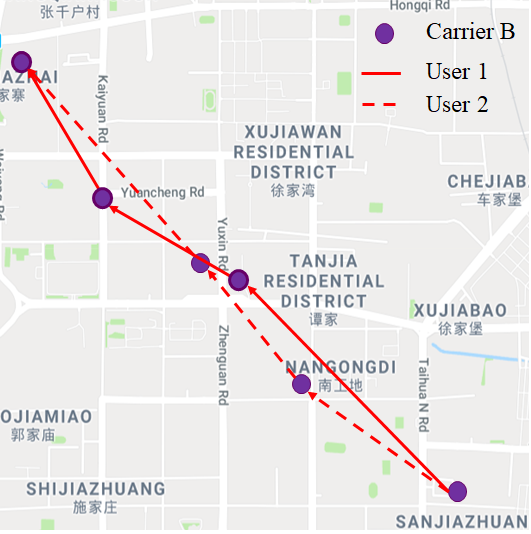}}
	\subfloat[Different carriers]{
		\label{figCellTowerLocationsDifferent}
		\includegraphics[width=1.5in,height=1.3in]{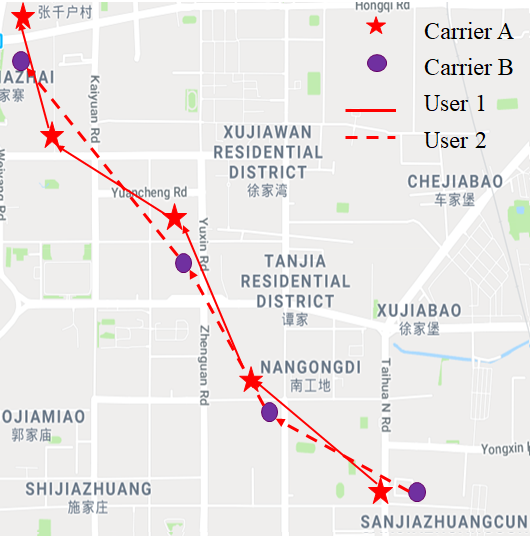}}
	\vspace{-0.1in}
	\caption{Cell tower sequences of two persons that move together.}
	\vspace{-0.2in}
	\label{figCellTowerLocations} %% label for entire figure
\end{figure}

Fig.~\ref{figSimilarityExample} depicts an example of the search process of a typical algorithm, i.e., Dynamic Time Warping (DTW)~\cite{dtw}. The query trajectory $Q$ is highlighted by the solid line and the dataset contains three trajectories, $\mathcal{T}=\{T_1, T_2, T_3\}$. 
First, DTW extracts a set of trajectories $\mathcal{T}'=\{T_1,T_2\}$ whose start-stop times are close to the query trajectory $Q$. Then, the similarity $s(T_i)$ between the query trajectory $Q$ and each trajectory $T_i$ in $\mathcal{T}'$ is calculated. In this example, $s(T_1)$=0.86 and $s(T_2)$=0.82. Finally, the similar trajectories ($T_1$) are retrieved if their similarity is
larger than the threshold (0.85).
% A set of new trajectory similarity measures have been developed for more efficient search~\cite{Chen2009Design,xie2017distributed,erp}.

%\begin{equation}
%Sim_{ST}(\tau_1, \tau_2) = \lambda \cdot Sim_S(\tau_1, \tau_2) + (1 - \lambda) \cdot Sim_T(\tau_1, \tau_2)
%\end{equation}

%distance equation and explanation
%Modify the figure or the above example, if they do not match.

\vspace{-0.1in}
\subsection{Cellular data}
\label{cellular}
In this study, we use an anonymized cellular dataset from a large city. 
The dataset contains the records over 61 days from two major mobile carriers, i.e. Carrier A and Carrier B. 
They provide 71.7\% and 28.3\% data in the dataset respectively. 
Table~\ref{table1} describes the format of each record, including Anonymized ID, Time, LAC (Location Area Code), and CID (Cell Tower ID). 
To protect the privacy of users, the carriers anonymize the data by replacing the subscriber identifications by a hash code.
The data do not contain any information relating to text messages, phone conversations or data usage. 
Based on the cell tower map provided by the carriers, we know the physical location of each cell tower. Thus, in our dataset, a cell tower sequence is a vector composed of the locations of all cell towers accessed by a mobile phone at different timestamps. 

\begin{table}[htbp]
	\caption{Examples of the records in our cellular dataset.}
	\vspace{-0.1in}
	\small
	\label{table1}
	\setlength{\tabcolsep}{5mm}{
		\begin{tabular}{llllll} % 控制表格的格式
			\hline 
			ID & Time  & LAC & CID \\
			\hline 
			1B2A7 &  20170901080234 & 37146 & 196*18  \\
			5U2F1 & 20170903070821 & 37149 & 195*57 \\
			\hline
	\end{tabular}}
\end{table}

%Due to the different usage of mobile apps, different persons may have different sample rates of their cellular data. The deviation of the average sample rates is up~to~0.0078 sample per second (corresponding to a sample interval of 128 seconds) between different subscribers.

\textbf{Data coverage.}
Our dataset covers 83.5\% of the total population in a big city. By filtering out the cell tower sequences of extremely-low sample rates (i.e., less than 1 sample per 10 minutes), we still have one third of records of total population. 
Fig.~\ref{fig2Xian} presents the density of cell towers and mobile users provided by the cellular dataset. 
The analysis results show that cellular networks can cover the city with a high density of cell towers and users for both urban and rural areas. 
% However, we also note that the density of cell towers and users varies in space. Special design are needed to handle this challenges. 
% Besides, based on our analysis on the 61-day cellular data, the density of mobile users does not vary much for different days.

%Figure \ref{figDensityUsers} depicts the density of mobile users in the city. The color represents the number of users within 1 $km^2$ during one day (Monday). Based on our experiments, the density of mobile users does not vary much for different days.

%Figure \ref{figCDFSampleRate} depicts the Cumulative Distribution Function (CDF) distribution of the average sample rates over one day for all subscribers. 

\vspace{-0.1in}
\subsection{Comparisons of cell tower sequences.} 
We exploit one question: \textit{can we detect two persons that move together by directly comparing their cell tower sequences?} 
Fig.~\ref{figCellTowerLocations} presents an example using the cell tower sequences of two persons that move together under two cases, i.e. they subscribe to the same carrier and different carriers.

\textit{Same carrier.} Even with the same carrier, the mobile phones of two persons that move together may associate with different cell towers. For example, in Fig.~\ref{figCellTowerLocations}~\subref{figCellTowerLocationsSame}, the sample points of the user 1 are different from that of the user 2, causing significant misalignment of the timestamps. It may be caused by the diverse app usage of different persons and the handover algorithm of the carrier.
 
\textit{Two different carriers.} 
Fig.~\ref{figCellTowerLocations}~\subref{figCellTowerLocationsDifferent} demonstrates that two persons that move together associate with completely different cell towers of different carriers. 
From the cell tower sequences, it is difficult to infer that the two persons are moving together.

The above two observations indicate that we cannot directly compare two cell tower sequences. 
First, the low sample rates of cellular data make aligning two cell tower sequences along the temporal dimension rather difficult.
Second, it is hard to find an appropriate distance threshold to determine whether two cell tower sequences are made by two similar moving trajectories, due to the location ambiguity of cellular data.
% These motivate us to investigate a novel method which can effectively search similar trajectories of multiple carriers. 

%Therefore, we design an architecture cellSim, which first match the cell tower sequence on a same road network to mitigate large difference caused by multiple carriers and improve the location accuracy.
%Then, we organically incorporate map matching results into a novel trajectory similarity search, which allows the accuracy alignment between the query trajectory and the candidate trajectories. 

\vspace{-0.1in}
\section{Design of cellSim}
% Due to the ambiguity caused by cell tower locations and multiple carriers, we cannot directly compare the raw cell tower sequences. 
% Therefore, we propose a system, cellSim, which performs a cross-layer design to organically incorporate map matching to trajectory similarity search. 
% We first match the cell tower sequence on a same road network to mitigate large difference caused by multiple carriers and improve the location accuracy.
% Then, we incorporate map matching results into a novel trajectory similarity search, which allows the spatio-temporal alignment between a query trajectory and the trajectory candidates. 
%We first transform the cell tower sequences into most likely trajectories on a same road network and process the generated trajectories by similarity search methods. 
We introduce an overview of cellSim and the design of two key components in cellSim, i.e., map matching and similarity search. Table \ref{notation} lists the notations used in this study.

\vspace{-0.1in}
\subsection{Overview}
\begin{figure}[t]
	\centerline{\includegraphics[width=.46\textwidth]{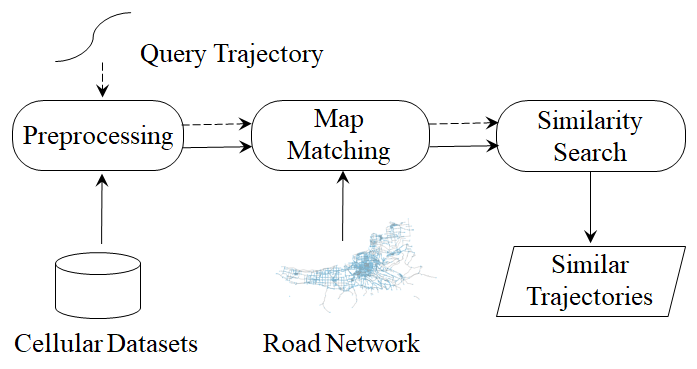}}
	\vspace{-0.1in}
	\caption{ The architecture of cellSim.}
	\vspace{-0.1in}
	\label{figArchitecure}
\end{figure}

Fig.~\ref{figArchitecure} depicts the architecture of cellSim. 
We first preprocess the cell tower sequences of our raw cellular dataset to mitigate noise and abnormal behaviors (Section \ref{secPreprocessing}). 
Each processed cell tower sequence is then transformed into a trajectory on a road network using a HMM-based map matching method (Section \ref{mapmatching}). 
The map matching processing of the cellular dataset can be conducted offline, without impacting the speed of our online similar trajectory search.
The output of map matching is a trajectory dataset generated from the cell tower sequence dataset.

The input of cellSim is a query trajectory, which can be either a cell tower sequence or a GPS trajectory on the road network. For a cell tower sequence, it is first transformed to a trajectory by the map matching component.
Given the query trajectory, the trajectory similarity search component calculates its similarity measure with each trajectory generated by our large-scale cellular data (Section~\ref{secTrajectorySimilarity}).
The output of cellSim is the similar trajectories which are close to the query trajectory in the space-time domain with a similarity larger than a threshold $\tau$.

\begin{table}[t]
	\caption{Notations used in this paper.}
	\vspace{-0.1in}
	\label{notation}
	\small
	\renewcommand{\arraystretch}{1.1}
	\setlength{\tabcolsep}{4.5mm}{
	\begin{tabular}{cp{4.8cm}} \hline 
		\textbf{Notation}  &  \textbf{Description} \\ \hline
		$p_i$  & A sample point in cell tower sequences  \\
		$T=p_1, p_2, ..., p_n$  & A trajectory consist of $n$ points \\
		$Len(T)$  & The distance length of trajectory $T$  \\
		$\mathcal{T}$  & The trajectory dataset\\
		$Q=q_1, q_2, ..., q_m$  & A query trajectory consists of $m$ points  \\
		$M$  & The number of trajectory candidates generated for one cell tower sequence\\
		$T^{'}$  & Top $M$ trajectory candidates for trajectory $T$ \\
		$d(p_{i-1},p_i)$  & The geodesic distance between the sample point $p_{i-1}$ and $p_i$  \\
		$T_{i-1,i}$  & A segment of $T$ between $p_{i-1}$ and $p_i$\\
		$c_i^j$  & The $j_{th}$ road segment of $p_i$ \\
		$d'(c_{i-1}^j, c_i^k)$  & The distance between the road segment $c_{i-1}^j$ and $c_i^k$ \\
		$L(Q,T)$ & The length of overlapping parts between $Q$ and $T$ \\
		$\tau$ & The similarity threshold\\
		% $SIM(Q,T)$ & The similarity of $Q$ and $T$\\
		\hline 
	\end{tabular}}
	\vspace{-0.1in}
\end{table}

\textbf{Holistic design.} 
Our map matching component outputs top $M$ trajectory candidates for each cell tower sequence (Section~\ref{subsubSecMultiple}).
Every trajectory candidate is also assigned a confidence to quantify its probability of being the true trajectory.
% It is more likely that the candidate set includes the true moving trajectory experienced by the user.
To measure the similarity of two cell tower sequences, our similarity search component performs pairwise comparisons between the $M$ trajectories of the query cell tower sequence and the $M$ trajectories of each cell tower sequence in the dataset, resulting in $M^2$ similarity comparisons. We select the highest similarity as the final similarity result for each cell tower sequence in the dataset. 
CellSim identifies all cell tower sequences with a similarity higher than a threshold $\tau$ as its final result.
To minimize false positive, $M$ is adapted according to the length of query trajectory (Section \ref{subsubSecAdaptive}).

Due to the expense of processing all sample points during calculating the similarity of $M^2$ trajectory pairs, we propose two pruning techniques, i.e., global pruning and local pruning, to skip some calculation without impacting the accuracy of our similarity search results~(Section~\ref{subsubSecPruning}).

\vspace{-0.1in}
\subsection{Map matching for cellular data}
\label{mapmatching}
We adopt a classic first order Hidden Markov Model~\cite{Newson2009Hidden} with full use of road network and cell tower information. 
Although many map matching algorithms have been proposed~\cite{Lafferty2001Conditional,Kundu1989Recognition},  they are not suitable for our scenario. 
For example, Conditional Random Field (CRF) models can model high-order dependencies among multiple states \cite{Lafferty2001Conditional}.
A lightweight map matching method~\cite{CRF} applies CRF for indoor localization. 
However, CRF needs to estimate a set of parameters beforehand. It is difficult to do so in our case, because we do not have sufficient training data at city scale.  
The other existing map matching algorithms, e.g., second order HMM~\cite{Kundu1989Recognition} or particle filters~ \cite{Rai2012Zee}, suffer from high time complexity.
The experimental results show that the calculation time of second order HMM and particle filters are more than three times that of first order HMM, due to the fact that they require more computation to evaluate the probability distribution. This makes it impossible to match massive trajectories.

\vspace{-0.1in}
\subsubsection{Customized HMM-based map matching}
\label{subsubHMM}
For the input of map matching, we have a time-series sequence of cell tower locations and the road segment information, including start and end points on the map, speed limit, etc. 
Our map matching model uses the input to find a trajectory composed of several road segments that are traversed by the user.
The HMM model has a hidden state and an observable state at each time step. Both states maintain two probabilities (emission and transition probability) to evaluate whether the user is at a specific road segment. 
% The probability of a moving trajectory being the true trajectory of the user is determined by the above two probabilities of each road segment on that trajectories. 
At the beginning, all road segments are initialized with the same probability. 
As the modeling process proceeds, some road segments' probabilities increase faster than the others. 
Finally, one trajectory composed of several road segments with high probabilities will be selected. 

Based on the general HMM process, we take further information of specific road networks and cell towers into account. Two customized improvements are made to the emission probability and the transition probability respectively as follows.

\textbf{Emission probability.}
It is the probability of observing a cell tower if the user is actually on the road segment~$j$.
As previous works have done \cite{Newson2009Hidden,aiji}, we model the emission probability of a projected point on the road segment as a Gaussian distribution given an observation point $p_i$ (cell tower). 
For an observed cell tower, it is more likely that the user is on some closer roads. 
For one state $c_i^j$ (i.e. $j_{th}$ road segment of $p_i$) and an observation $p_i$, the emission probability is given as:
\begin{equation}
P(p_i,c_i^j)= \frac{1}{\sqrt{2\pi}\sigma_t} e^{-\frac{1}{2}(\frac{w_{d}*w_s*d(p_i,c_i^j)}{\sigma_t})^2}
\end{equation}
where $\sigma_t$, the standard deviation, is adapted to the cell tower density, $d(p_i,c_i^j)$ is the distance between $p_i$ and $c_i^j$, and $w_d$ and $w_s$ are two weight parameters derived by the hints from the road network. 
They are based on the following two observations.

First, \textit{people normally prefer to follow the same direction or only slightly deviate from the moving direction}. 
For example, on a two-way road, people are more likely to keep the same forward direction, rather than changing to the opposite side of the road. 
%The observation is confirmed by \cite{assumption1}. Mondal~\textit{et. al.} first selected six typical study areas with different traffic conditions in a city. Then, they presented details of traffic characteristics at the selected areas by extracting the number of vehicles in through traffic and U-turns using the installed video recorders. The result showed that on average 93.4\% of the persons preferred straight road.
To map the road direction to a weight $w_d$, we use the following function.
\begin{equation}
\vspace{-0.05in}
w_{d} =|dir_s-dir_c|/2 \pi
\vspace{-0.05in}
\end{equation}
where $dir_s$ is the moving object's direction vector, and $dir_c$ is the candidate road segment's direction vector.

Second, \textit{people normally prefer to use a wider road, if multiple roads can lead to the destination}. 
%This observation is verified by transportation statistics~\cite{british}. These statistics demonstrate that motor vehicles travelled 326.2 billion miles from April 2017 to March 2018 and 65.7\% of the total miles on wider roads (i.e., 'A' roads and motorways).
We assign the roads of a higher speed limit with a larger weight in their observation probability, as follows:
\begin{equation}\label{cspeed}
\vspace{-0.05in}
w_s = 1 - c \cdot r_l
\vspace{-0.05in}
\end{equation}
where $c$ is constant and $r_l$ is the speed limit of roads with a maximum speed limit of 120 km/h.

%\begin{figure}[tbp]
%	\centerline{\includegraphics[width=3in]{figs/figExampleBidirectional.png}}
%	\caption{Example of Bidirectional}
%	\label{figExampleBidirectional}
%\end{figure}
%Figure \ref{figExampleBidirectional}

\textbf{Transition Probability.}
The transition probability measures the probability that the user transits from a state $j$ at step $i-1$ to another state $k$ at next step $i$. 
Most previous works use the idea proposed by Newson and Krumm \cite{Newson2009Hidden}, i.e. the geodesic distance between two states (i.e. road segments) should be similar to the distance between two observations (i.e. cell towers). 
The transition probability that the user moves from one road segment $c_{i-1}^j$ to another road segment $c_i^k$ can be expressed as follows.
\begin{equation}\label{equTransition}
P(c_{i-1}^j,c_i^k)=\frac{1}{\beta} e^{-D(c_{i-1}^j,c_i^k)/\beta}
\end{equation}
where $\beta$ is a constant parameter, experimentally set to 0.0096 in our implementation. $D(c_{i-1}^j,c_i^k)$ is the difference between the geodesic distance of two cell towers and the geodesic distance of two road segments. The transition probability is high if $D(c_{i-1}^j,c_i^k)$ is small.

The above method works for GPS trajectories with high sample rates. 
However, due to the low sample rate of our cellular data, people may travel for a long distance between two consecutive time points, and the geodesic distance between two cell towers cannot reflect the real distance traveled on the road map.
To consider all possible paths between two states (i.e. road segments), we argue that \textit{people are more likely to take the shortest path among all possible paths on the road network}.
%This observation is supported by human behavior studies~\cite{Zhu2015Do}. 
%In our dataset, the average distance between two adjacent sample points is about 412 meters. Based on the study in~\cite{Zhu2015Do}, with such a distance, more than 80\% of people follow the shortest path. 
%Therefore, 
$D(c_{i-1}^j,c_i^k)$ used in Eq.~\ref{equTransition} can be expressed as follows.
\begin{equation}
D(c_{i-1}^j,c_i^k) = |d'(c_{i-1}^j,c_i^k) - min_{c_{i-1}^m \in c_{i-1} , c_{i}^n \in c_{i}} \{d'(c_{i-1}^m,c_i^n) \}|
\end{equation}
where $d'(c_{i-1}^j,c_i^k)$ is the distance between two road segments $j$ and $k$ on the road network and $min_{c_{i-1}^m \in c_{i-1} , c_{i}^n \in c_{i}} \{d'(c_{i-1}^m,c_i^n) \}$ is the minimum distance of all possible paths at two consecutive time points. At time $i$, the hidden state ($c_{i}^n$) can be any road segment around the cell tower ($c_{i}^n \in c_i $).   

%\begin{figure}[tbp]
%	\centerline{\includegraphics[width=2in]{figs/figExampleTransitionMatrix.png}}
%	\caption{Example of Transition Probability}
%	\label{figExampleTransitionMatrix}
%\end{figure}

\begin{figure}[tbp]
	\centerline{\includegraphics[width=2.8in,height=1.5in]{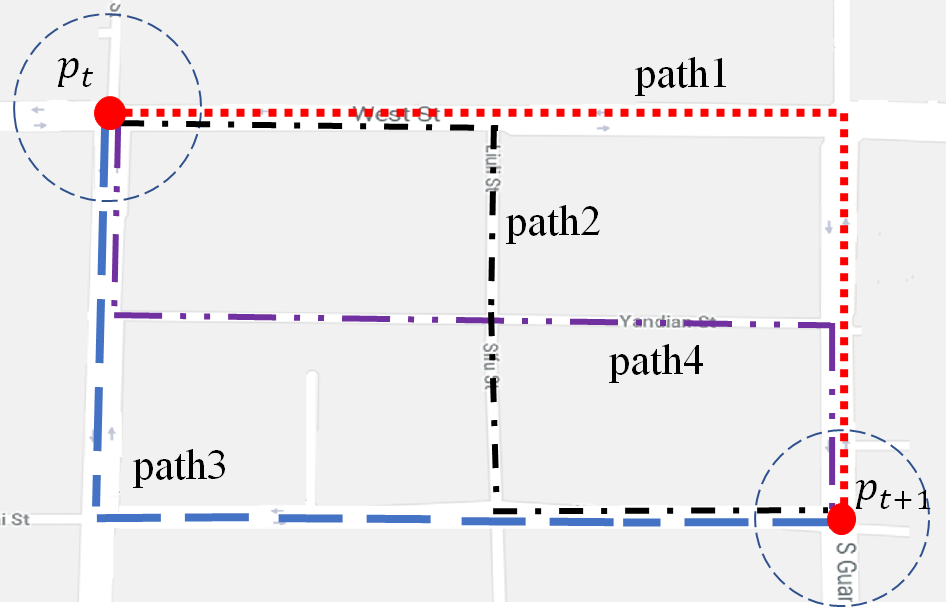}}
	\caption{Example of multiple paths.}
	\vspace{-0.2in}
	\label{figMultipleResults}
\end{figure}

\vspace{-0.1in}
\subsubsection{Multiple trajectory candidates}
\label{subsubSecMultiple}
Due to the low sample rate, there may exist several possible paths with similar distances from one cell tower to another one.
It is hard for the HMM transition probability to distinguish these paths. 
As the example in Fig.~\ref{figMultipleResults} illustrates, for cell tower $p_{i-1}$ and $p_{i}$, if their candidate road segments are the closest intersections (highlighted as red points in the figure), there are 4 similar paths. 
To handle this problem, we output multiple trajectory candidates as the results of map matching. We will use the information from the other time points in the whole journey to find the best trajectory during the similarity search process.

We do not need to output multiple paths between two cell towers when their projected two road segments only have one path that connects them. We detect such a case if the length of the path is equal to the geodesic distance between two road segments.
One example is a long straight road on which the cell towers are projected.
The cell towers are located along with the road, and $d'(c_{i-1}^j, c_i^k)$ and $d(p_{i-1},p_i)$ are similar.
In this case, we only output one path which has the highest probability of HMM model as the result.

We can now find multiple paths between two cell towers. 
For a sequence of many cell towers, we finally output top $M$ trajectory candidates, each of which is composed of the paths from multiple pairs of cell towers. 
We first perform the map matching process to obtain one trajectory with the highest HMM probability. 
Then, we use the above method to verify whether it is necessary to output multiple paths for each pair of adjacent cell towers along the sequence one by one.
If necessary, we generate $M$ paths and replace the original path in the trajectory. 
By splicing the multiple paths of all pairs, we obtain a number of trajectory candidates that are more than $M$. 
Suppose a trajectory has $m$ samples, we calculate its confidence in its map matching result $P_{map}$ by multiplying the emission probability of each road segment by transition probability between the road segments as follows.
\begin{equation}\label{pm}
\vspace{-0.1in}
P_{map} = \prod_{i=1}^{m} P(p_i,c_i) \times P(c_{i-1},c_i)
\end{equation}

Finally, we rank these trajectory candidates based on their normalized probabilities and output the top $M$ trajectories $T^{'}$. 

\vspace{-0.1in}
\subsection{Trajectory similarity search}
\label{secTrajectorySimilarity}
% Given a cell sequence $T$, we perform pairwise comparisons between its $M$ trajectory candidates and the $M$ trajectory candidates of each cell tower sequence in our dataset. 
% We repeat such $M^2$ comparisons for each cell tower sequence in our dataset. 
% First, a new trajectory similarity measure is proposed to calculate the similarity between two trajectories.  
% Second, the parameter $M$ needs to be carefully set to achieve a proper balance between false positive and true positive. 
% Finally, two pruning schemes are developed to reduce the computation overhead. 
In this section, we introduce our new similarity measure function, the adaptation algorithm of the parameter $M$ and the pruning schemes in sequence.

\vspace{-0.1in}
\subsubsection{A new trajectory similarity measure}
\label{simi}
\begin{figure}[tbp]
	\centerline{\includegraphics[width=2.8in,height=1.2in]{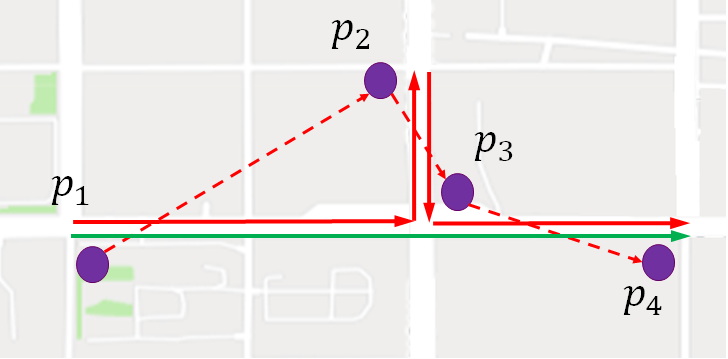}}
	\caption{Example of ambiguity caused by cellular data.}
    \vspace{-0.2in}
	\label{noise}
\end{figure}

Although the above map matching scheme can significantly mitigate the location ambiguity of cellular data, the output trajectories are still not perfect. Fig.~\ref{noise} depicts an example of four sample points. 
Our map matching algorithm outputs the result as a solid red line.
Compared with the real trajectory (dashed green line), the map matching result has an extra u-turn caused by the large localization error of $p_2$ and $p_3$. 
We have handled this problem by assigning a larger weight for keeping the same direction in map matching, due to the low localization granularity of cellular data.

%我们提出的轨迹相似度方法 怎么做的
The above problem results in the low performance of existing methods for processing our cellular data, since they are based on the distance between the query trajectory and a trajectory candidate. Sudden changes of trajectory shapes can cause the similarity distance to increase sharply.
We investigate a novel trajectory similarity measure to handle these sudden changes of the trajectory shapes in the process of similarity search. 
Our measure calculates the overlapping ratio between the query trajectory and a trajectory candidate.
It minimizes the impact of the deviated parts in trajectories, which increases its ability to tolerate noise.
% At the same time, if two or more people traverse on the same road, the overlapping ratio of their trajectories can better reflect whether they are co-moving.

% 对待比轨迹进行时间插值 方便时间窗构造
To compare a query trajectory $Q$ with a trajectory candidate in the dataset $T$, we first align them in time by the temporal interpolation on road maps. 
We use the timestamps of the original cell tower sequence to calculate the moving speed by the distance on the road network and time differences of two consecutive projected points. 
Then, we estimate the time information of any point $p_i$ between these two projected points by the movement distance and moving speed.

With two aligned trajectories, we use a sliding window to extract a segment of these two trajectories separately. For a sample point $q_i$ in the query trajectory $Q$, there must exist a point $p_i$ in the trajectory candidate $T$ that has the same timestamp with $q_i$.
For the segments $Q_{i-1,i}$ and $T_{i-1,i}$, we calculate the length of their overlapping parts, $L(Q_{i-1,i},T_{i-1,i})$. 
To calculate the similarity between $Q$ and $T$, we move the sliding window forward and obtain the total length of the overlapping part in all sliding windows. 
The final similarity is the overlapping ratio, the total length of the overlapping parts normalized by the length of query trajectory. 
For a query trajectory $Q$ and a trajectory candidate $T$, we also consider their map matching confidences $P^Q_{map}$ and $P^T_{map}$. Supposing the query trajectory $Q$ has $m$ samples ($Q=Q_{1,m}$), its similarity to the trajectory $T$ can be calculated~as: 
\begin{equation}
\label{pmap}
Sim(Q, T) = P^Q_{map} \times P^T_{map} \times \frac{\sum_{i=1}^{m} {L(Q_{i-1,i},T_{i-1,i})}}{Len(Q)}
\end{equation}

\vspace{-0.1in}
\subsubsection{Adaptive similarity search}
\label{subsubSecAdaptive}
By outputting multiple trajectory candidates with different map matching probabilities, we increase the probability of including the true trajectories in the $M^2$ comparisons.
Moreover, if two persons move together, their trajectories are almost identical and should have the highest similarity, compared with the other $M^2-1$ pairs of trajectories.
As a result, the probability of successfully detecting two co-moving persons is increased. 
For example, if the cell tower sequence of another user has more samples between $p_{i-1}$ to $p_{i}$ in the example of Fig.~\ref{figMultipleResults}, the information of extra cell towers helps HMM model to find one path with the highest probability. 
The results of four pairwise comparisons are likely to include the comparison between two true paths. 
On the other hand, multiple trajectory candidates may also increase the false positive where cellSim identifies two similar trajectories, which are not in fact, from two  persons that move together.

For long query trajectories, the false positive problem is rare. The example shown in Fig.~\ref{figMultipleResults} only relates to the paths between two cell towers, corresponding to a short part of the whole trajectory. If two persons do not move together, the remaining parts of their trajectories will produce a low similarity result. 
However, when the query trajectory is short, multiple trajectory candidates may increase this false positive. 

To minimize false positive, we adapt the number of trajectory candidates used in the similarity search according to the length of the query trajectory. We categorize the trajectory length into 5 levels, i.e., \{$<3km$\}, \{$\geq3km \ \& <6km$\}, \{$\geq6km \ \& <9km$\}, \{$\geq9km \ \& <12km$\} and \{$\geq12km \ \& <15km$\}. For each level, we find the best setting for the number of trajectory candidates (i.e. $M$), which is proportional to the trajectory length (see the empirical experiment results in Section \ref{len}).

\vspace{-0.1in}
\subsubsection{Pruning}
\label{subsubSecPruning}
Although the similarity comparisons can be executed in parallel to speed up the similarity search of cellSim, we propose two orthogonal pruning techniques to reduce the computation overhead of cellSim.

\textbf{Global pruning.}  
For one query trajectory, we do not need to evaluate every cell tower sequence in the dataset. 
Instead, we propose a global pruning method, which skips a large portion of sequences whose start and end points are significantly different from those of the query trajectory. 
A pair of points $p_i(x_i,y_i,t_i)$ and $p_j(x_j,y_j,t_j)$ are close with each other if they meet the following requirement.
\begin{equation}\label{glopruning}
\sqrt{| x_i-x_j | ^2 + | y_i-y_j | ^2 + | t_i-t_j | ^2} \leq \epsilon
\end{equation}
where $\epsilon$ is the global threshold and set dynamically according to the density of cell-towers.
If the density of cell-towers is small, a relatively-larger threshold $\epsilon$ is used.

\textbf{Local pruning.}  
For a single cell tower sequence, we devise a local pruning method to avoid some unnecessary comparisons in the $M^2$ comparisons. 
%We need to compare a query trajectory $Q$ and the candidate trajectories of one cell sequence ($\mathscr{T} = \{T_1, T_2, ..., T_M \}$).  
%First, we sort all candidate trajectories by their confidences of the map matching result in descending order. Suppose the sorted result is $\mathscr{T} = \{T_1, T_2, ..., T_M \}$.  
%First, we calculate the similarity $s_i$ between $Q$ and $T_i$ using the similarity measure. 
%we define a lower bound $\tau$. 
Our similarity measure (see Section \ref{simi}) is computed step by step. At each step $n$, we calculate the overlapping length $L(Q_{n-1,n},T_{n-1,n})$ and the accumulated similarity.
%\begin{equation}
%Sim(Q_{1,n}, T_{1,n}) = P_{map} \times \frac{\sum_{i=1}^{n} {L(Q_{i-1,i},T_{i-1,i})}}{Len(Q_{1,n})}
%\end{equation}
According to Eq.~\ref{pmap}, the accumulated similarity is decreasing as the step index $n$ increases. During the computation of the similarity for one trajectory~$T_i$, we stop the computation and $T_i$ is pruned, if its accumulated similarity $Sim(Q_{1,n}, T_{1,n})$ is lower than the similarity threshold $\tau$. 

\vspace{-0.1in}
\subsection{Preprocessing of cellular data}
\label{secPreprocessing}
We develop three noise filters to process noisy phenomena in our cellular data before map matching.

\textbf{Ping-Pong filters.}
When a mobile user is in the middle of several adjacent cell towers, her phone performs handover between several cell towers frequently in a short time.
For each sample point, our filter first checks whether the following $w_p$ sample points have the Ping-Pong phenomenon, i.e., some points are from cell towers that differ from the current point, while others are from the same cell tower as the current point. If a Ping-Pong phenomenon is detected, the sample points of the second cell tower are discarded. 
Sometimes, we cannot handle all the Ping-Pong noises at once. We apply the Ping-Pong filter to process the same cell tower sequence iteratively until the sequence size does not decrease. 

\textbf{Backward filters.} 
When a person moves in a direction, the mobile phone may sometimes handover to a cell tower opposite to the moving direction due to the fluctuation of wireless signals. 
As a consequence, the cell tower sequence may change direction suddenly at some points and return to the normal direction shortly. 
This phenomenon causes U-turns in the map matching results. 
To handle this noise, when the moving direction of one point is different from the current direction, the filter caches that point and checks its next $w_b$ points. If the next $w_b$ points confirm the change of direction, the filter adds all points to the result; otherwise, it deletes that cached point and keeps the $w_b$ points.

\textbf{Drifting filters.}
A mobile phone may sometimes handover to a cell tower very far from its location. As a consequence, its cell tower footprint sometimes moves quickly to another location, and then returns to the previous location with an impractical moving speed. 
To handle the drifting noise, we estimate the moving speed between two points, based on their distance and time interval. If the moving speed is higher than the road speed limit, the next point will be discarded.

\vspace{-0.1in}
\subsection{Discussion}
\textbf{CellSim Assumptions.}
We shows the validity and rationality of the assumptions in cellSim.

First, we assume that people normally prefer to follow the same direction or slightly depart from the moving direction. The assumption is confirmed by~\cite{assumption1}. 
In that work, the authors first selected six typical study areas with different traffic conditions in a city. Then, they presented details of traffic characteristics at these areas by extracting the number of vehicles in through traffic and U-turns using installed video recorders. The result showed that on average 93.4\% of the persons preferred 
straight road.

Second, we assume that people normally prefer to use a wider road, if multiple roads can lead to the destination. This assumption is verified by transportation statistics released by Britain in 19th July 2018~\cite{british}. The statistics showed that motor vehicles travelled 326.2 billion miles from April 2017 to March 2018 and 65.7\% of the total miles is moving on the wider roads (i.e., 'A' roads and motorways).

Third, we assume that people are more likely to take the
shortest path among all possible paths on the road network. Zhu \textit{et al.} \cite{Zhu2015Do} revealed that the larger the distance between origin and destination, the less likely people take the shortest path. 
In our dataset, the average distance between two consecutive points is about 412 meters, which means more than 80\% of persons will follow the shortest path. 

%we need to combine the cell tower distribution map with the road network map.
%On the road network, each road is divided into several straight segments, i.e., the average length of one road segment is 423.63 meters. Each road segment is recorded by the coordinates of its two end points. The road network is formed by connecting all road segments with their end points.

\textbf{Road segment candidates of each cell tower.}
To minimize the computation of map matching, we set a limited searching range for each cell tower to find candidate road segments.
The searching range for each cell tower is dynamically determined by the local density of its surrounding cell towers.
We adopt a linear function to fit the relationship between local density of cell towers and the searching range.
When the density is larger than $50/km^2$, the searching range is set to 200 meters, and when the density is less than $5/km^2$, the searching range is set to 1000 meters.
All the road segments that fully or partially fall into the range of a cell tower are candidates for its map matching.

\textbf{Distance between a cell tower and a road segment.} For the emission probability in map matching (Section \ref{subsubHMM}), we need to calculate the distance between a cell tower and a road segment. This is calculated as the geodesic distance between the cell tower and its projected point on the road segment. If the projected point is on the extension of the road section, we use the close endpoint of the road segment as the projected point in the distance calculation. 

%we need to combine the cell tower distribution map with the road network map.
%On the road network, each road is divided into several straight segments, i.e., the average length of one road segment is 423.63 meters. Each road segment is recorded by the coordinates of its two end points. The road network is formed by connecting all road segments with their end points.

%\textbf{Implementation cost.}
%The main deployment cost of cellSim consists of two parts: execution cost and data storage cost. 
%As to execution cost, cellSim implements two plug-in components (i.e. map matching and trajectory similarity search). Map matching component is a costly operation which need a cluster with adequate nodes to process incoming data stream in real time. Compared with map matching, trajectory similarity search component has a relatively lower cost.
%Second, our dataset contains the mobile phone usage of 83.5\% of the total population in 61 days in a big city. The daily volumn is about 450 GB, which need the preprocessing and storage of real-time data streams  In our experiments,

\vspace{-0.1in}
\section{Implementation}
\label{implet}
We implement cellSim on a Hadoop cluster with 3 master nodes and 10 slave nodes. 
Each master node uses a dual Intel E5-2680 v4 CPU @ 2.4GHz with 14 cores. Each slave node has a dual Intel E5-2650 v4 CPU @ 2.4GHz with 12 cores. The total RAM is 1.5 TB and the total storage is 260 TB. The nodes run on CentOS release 6.8 with Hadoop 2.6.0-cdh5.5.0. 

\begin{figure}[tbp]
	\centerline{\includegraphics[width=3.5in, height=1.5in]{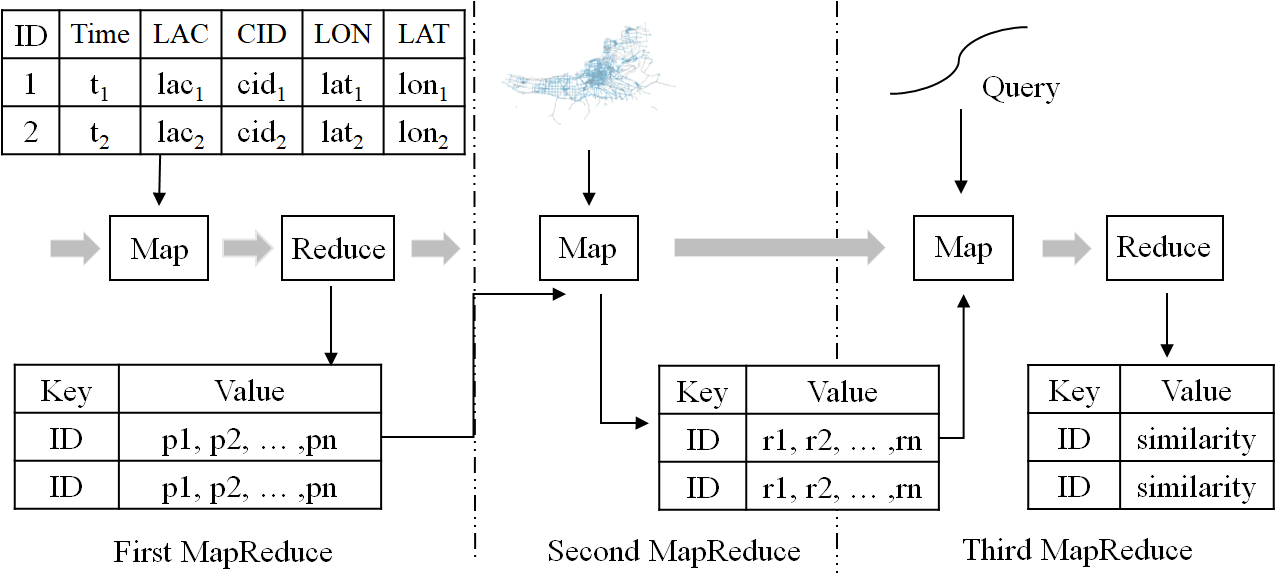}}
	\vspace{-0.1in}
	\caption{CellSim workflow with MapReduce.}
	\vspace{-0.2in}
	\label{figWorkFlow}
\end{figure}

\textbf{Implementation of cellSim based on MapReduce.}  
In our dataset, the daily data volume is $450$ GB. To process the data efficiently, we implement cellSim on MapReduce~\cite{hadoop} in  parallel.
%A distributed file system is deployed on a cluster of servers. 
Fig.~\ref{figWorkFlow} depicts the work flow of our framework that consists of three MapReduce jobs. 
%The first MapReduce job realizes the construction and preprocessing of discrete trajectories from cellular points for each anonymous user. In the second MapReduce job, the parallel map-matching algorithm are conducted to map discrete trajectories on to continuous road segments. In the final job, we retrieve trajectories that similar to query trajectory in the dataset based on the proposed similarity measure. The details of work flow are described below.

First, the raw cellular data are transformed into cell tower sequences. In the Map phase, the original logs are segmented by Hadoop TextSplitter. 
The logs from one user are identified by the same anonymized IDs and shuffled to the same reducer. In the Reducer phase, we sort all the records by time. 

Second, the map matching algorithm is implemented by a Mapper job. The digital road network and cell tower sequences are fed into the Mapper jobs. Through processing, multiple trajectory candidates are generated as a sequence of road segments with timestamps. 
% Each node stores a copy of the road network to improve the efficiency of data reading. 

Finally, we calculate the similarity between the query trajectory and every trajectory in the generated dataset. 
We first call the $setup$ function to read the query trajectory and match it with the road network. 
We compute the similarity using the trajectory similarity measure designed in Section~\ref{secTrajectorySimilarity} and output the trajectories with the similarity values larger than the similarity threshold.

The first two jobs can be done offline to process the cellular dataset. For the query of a cell tower sequence, we only need to convert it to $M$ trajectory candidates and perform the last similarity search job.

%\textbf{Reducer size adaption}  
%In the map-reduce model, the raw data are divided into pieces and computed by mappers in parallel. The number of mappers is normally set to the maximum according to the computing resource. 
%Since there are some operations, like shuffle and sort to categorize the trajectories into the same User-ID for reducer, that generate significant traffic and computing overhead, the setting on the number of reducers is important for efficiency. A good setting of numbers of reducer reaches a balance between the computing resource of the cluster and the balance of the load. We compute the reduce size according to the amount of data transferred from the map. When constructing user trajectory in first step, we need to transfer all user cellular logs to reducer and the reduce sizes needs to be increased. Thus, the reduce size is set to $1.75 * 10 \approx 18 $ . When computing user trajectory in third step, we need the similarity value and the number is reduced to 5 to support global ordering. 

%\textbf{Distribute cache.}  
%To provide distributed map matching, we need reserve road network on slave nodes. Traditional small data cache technology can not cache data of large map road nodes. To overcome this, we use distributed cache technology to realize the cache of road network. Distributed Cache is a tool provided by the MapReduce framework to cache files. It will copy the cache files on to the slave node before any tasks for the job are executed on that node. Its efficiency stems from the fact that the files are only copied once per job. 

\begin{figure}[tbp]
	\centerline{\includegraphics[width=3in,height=0.8in]{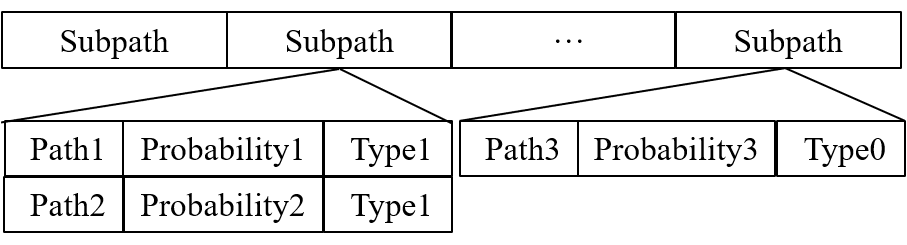}}
	\vspace{-0.1in}
	\caption{Data format in cellSim.}
	\vspace{-0.2in}
	\label{figDataFormat}
\end{figure}

\textbf{Data format.}  
For each cell tower sequence, we may need to store $M$ trajectories. If one trajectory occupies $B$ bytes, the total cost is $M*B$ bytes, which is memory-consuming. 
We develop a new data format to efficiently store all $M$ trajectories. 
% At the same time, the new data structure can also make our similarity search easier. 
As shown in Fig.~\ref{figDataFormat}, one trajectory is composed of multiple sub-paths. For all $M$ trajectories, some sub-paths are the same, e.g., some parts with straight road segments. In our data format, each sub-path has a binary field, ``Type'', which indicates whether this part is shared by all $M$ trajectories. If ``Type'' is 1, we only store one sub-path for all $M$ trajectories; otherwise, we provide a set of possible path instances. We also save the confidence of each sub-path according to the result of the map matching process. 

%If the $M$ trajectories are stored by rows, when they are read by the MapReduce tasks, they may be retrieved by different servers. After the calculation of similarity, the similarity results of $M$ trajectories should be gathered to one server and processed to output the final result. Our data format can enable all $M$ trajectories be processed on a same server. It is easier to process the results of all $M$ trajectories.

%\textbf{Global ordering.}  
%Achieving global ordering of similarity values may face a great challenge on efficiency, where similarities of all trajectory pairs are sent to one reducer for sorting. Meanwhile, it is likely to occur the problem of out of memory, especially when the data volume is large. To solve this problem, we implement a partition function to divide the similarity values into different segments and then pass all trajectories fallen in one segment into a reducer for global sorting.
%

\textbf{Implementation cost for different cities.}
The overhead of implementing cellSim for a new city is low. For different cities, we use different road maps and cellular datasets, which are implemented as two plug-in components in cellSim and can be changed easily without any modification of the other components. 
% The execution and storage of cellSim require adequate hardware resources. 
% To deploy our system for different cities, a cluster with adequate computational resources and storage space is needed.
Both the trajectory similarity calculation and the map matching modeling take the digital road network of the city as input.
The road network can be obtained from an open source websites (e.g. OpenStreetMap). The public road network data contain all the information we need in cellSim, including road speed limit, the start point and end point of each road segment, etc.

\vspace{-0.1in}
\section{Evaluation}
% We conduct extensive experiments to evaluate cellSim. 
We first present the experiment results on the overall performance of cellSim. 
Then, we study the performance of the two key components in cellSim, i.e., map matching and similarity search. 
Finally, we test the computation overhead of cellSim and the performance of the noise filters for map matching.  

\begin{figure*}[t]
\subfloat[Multiple carriers dataset]{
\includegraphics[width=0.3\linewidth,height=3.3cm]{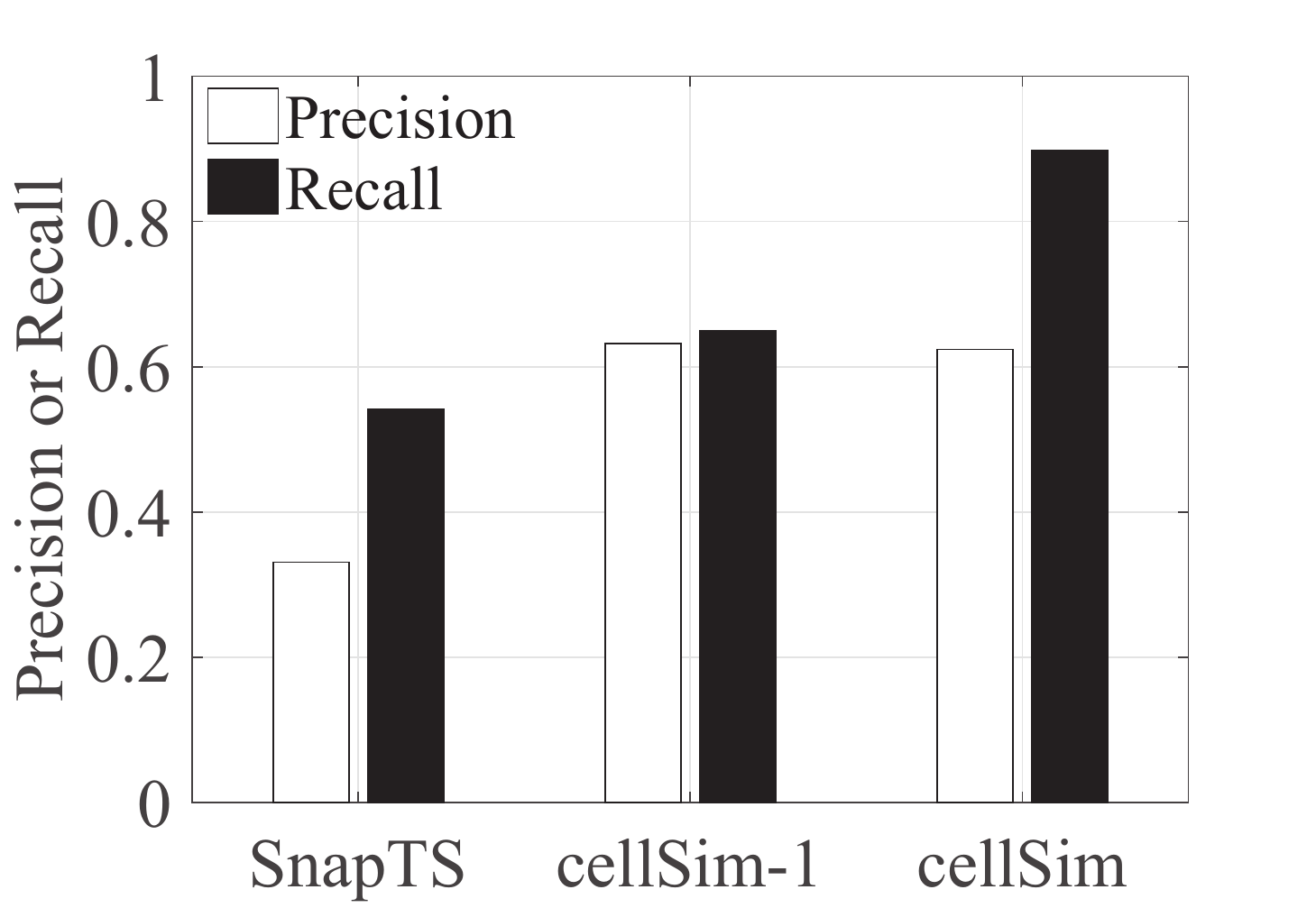}
\label{Cmulti}
}
\quad
\subfloat[Single carrier dataset]{
\includegraphics[width=0.3\linewidth,height=3.3cm]{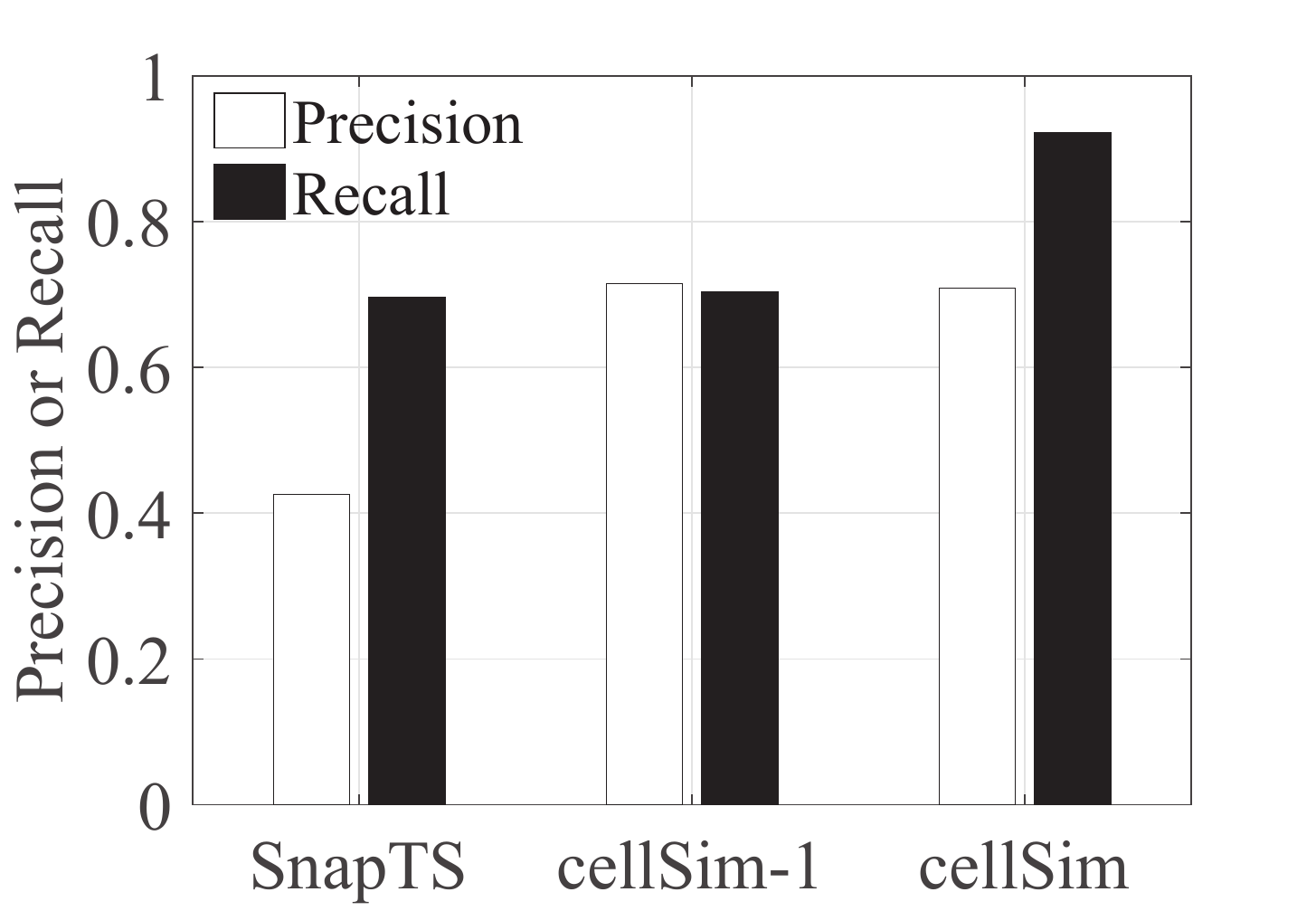}
\label{Csingle}
}
\quad
\subfloat[Query trajectory of GPS data]{
\includegraphics[width=0.3\linewidth,height=3.3cm]{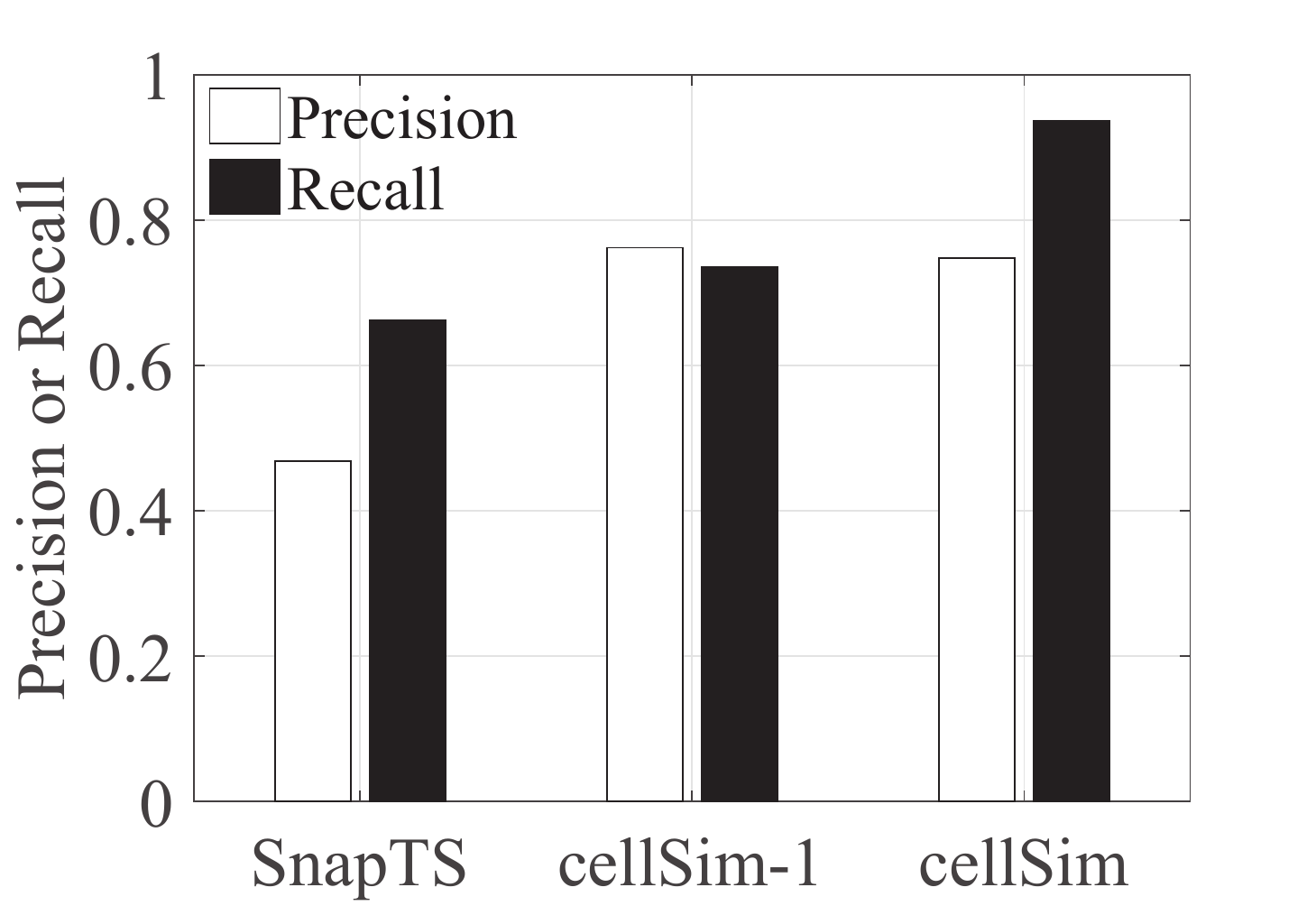}
\label{Cgps}
}
\vspace{-0.1in}
\caption{Overall performance of cellSim in terms of precision and recall on retrieving similar trajectories.}
\vspace{-0.2in}
\label{Ccomall}
\end{figure*}

\vspace{-0.1in}
\subsection{Experiment setting}
We test cellSim on the cellular dataset introduced in Section~\ref{cellular}. 

\textbf{Ground truth.} 
During the two months of our dataset, we also recruited 60 volunteers and collected their trajectories as ground truth to evaluate cellSim.
We assigned the volunteers to different groups, each of which contained 2-8 persons. 
Volunteers were required to enable the GPS on their mobile phones and record the location information at a sample rate up to 1 sample/sec, when they were moving outdoors.  
%Volunteers recorded their GPS trajectories and their co-moving persons, which are used as the ground truth in our evaluation. 
%The least and most number of co-moving persons are two and eight. 
At the same time, based on their phone number, our mobile carrier collaborators provided us their anonymized IDs which we used to identify their cell tower sequences in the anonymized cellular dataset.
All the volunteers have been informed with the purpose of their data collection and experimental procedure, and have signed the consent form.
In the end, we collected 1701-km trajectories that included 121 co-moving groups and covered both urban and rural areas with different road conditions, including main road and side roads. 
The GPS trajectories we used as ground truth in our evaluation experiments.
Our volunteer groups drove private vehicles to generate long-distance trajectories with varied speeds. Thus, it is unlikely that there are other people who may travel along with our volunteers.
For each co-moving group, one randomly-selected trajectory is used as the query trajectory to search for the other trajectories in the dataset.

\textbf{Baselines.}
Since cellSim is the first work searching for similar trajectories from cellular data, we build two baselines for evaluation.
First, to evaluate the design of adaptive $M$ trajectories in cellSim, we implement a simple version of cellSim, denoted as cellSim-1, in which only the top 1 trajectory candidate is generated by map matching.
Second, we develop a simple solution based on existing techniques. We replace the map matching component and the similarity search component of cellSim-1 by two state-of-the-art approaches: SnapNet~\cite{aiji} and TS-Join~\cite{vldb} respectively. The combined baseline is denoted as SnapTS.

We further compare the proposed map matching component of cellSim with two existing map matching methods. 
SnapNet~\cite{aiji} develops a HMM model that considers the road network information for cellular trajectories collected by mobile phones. 
ST-Matching~\cite{Lou2009Map} matches low sample rate GPS trajectories to the road network by combining both spatial and temporal features.

We also compare the proposed trajectory similarity measure of cellSim with two existing approaches. DTW~\cite{dtw} finds the optimal spatio-temporal alignment by matching one point of a trajectory with multiple points of another trajectory. 
TS-Join~\cite{vldb} calculates the spatial and temporal similarities separately. It then combines the similarity results of these two parts by weighted averaging with a parameter $\lambda$. In our implementation, $\lambda$ is set to~0.5, the best reported in~\cite{vldb}.

All the parameters of the above baselines are set to the optimal values that are obtained by empirical experiments on our cellular data, e.g., we set $\tau$ to 0.85 for both DTW and TS-Join to achieve their optimal performance.
% In the following experiments, without any additional notification, we conduct our experiment under the input of multiple carriers with adaptive parameter estimation of multiple candidates $M$.

\begin{figure}[tbp]
	\subfloat[Similarity threshold $\tau$]{\label{figTau} 
		\includegraphics[width=.23\textwidth,height=2.7cm]{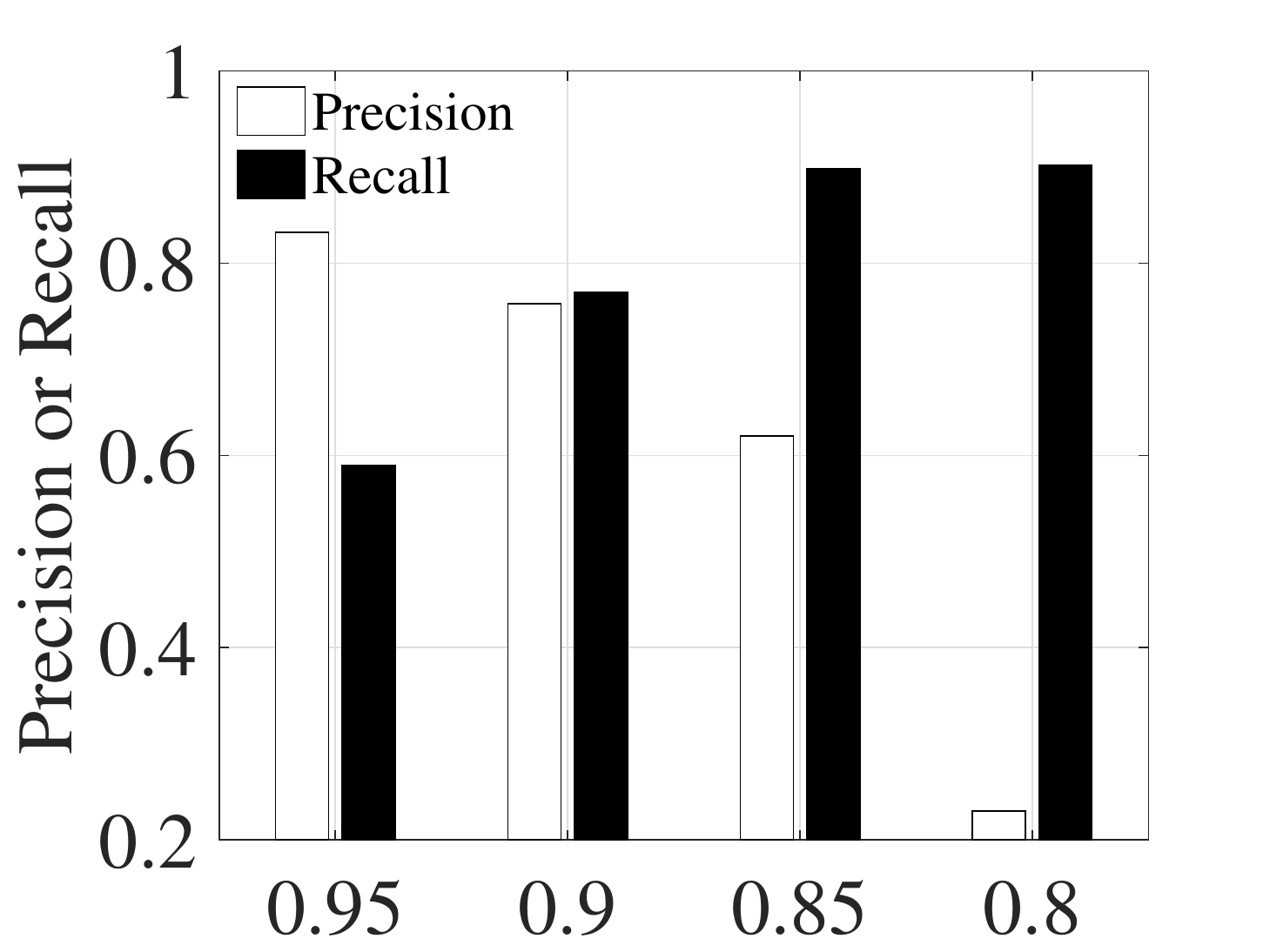}}
	\subfloat[Multiple trajectory candidates $M$]{\label{figM}
		\includegraphics[width=.23\textwidth,height=2.7cm]{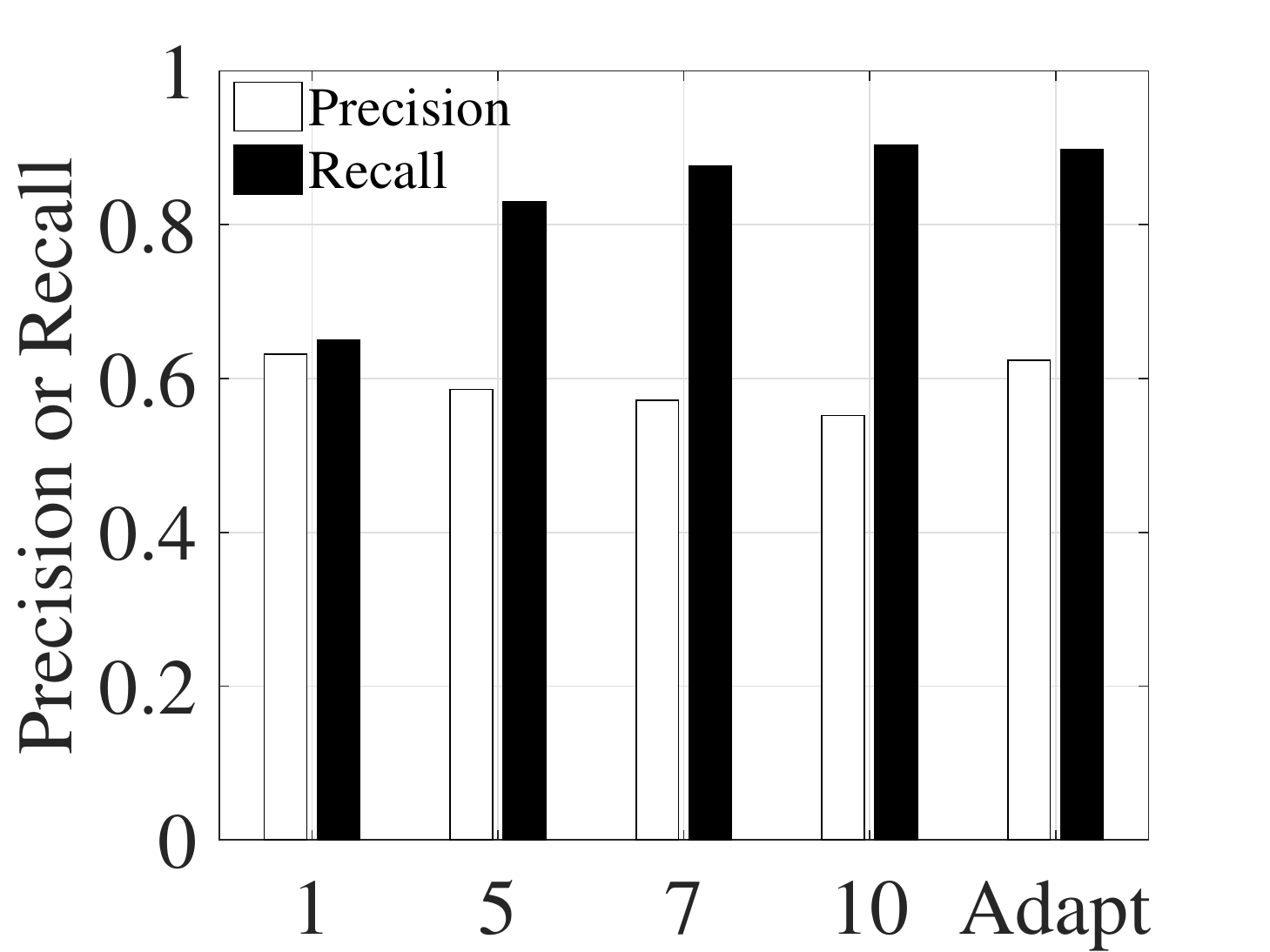}}
	\vspace{-0.1in}
	\caption{Effect of parameter $\tau$ and $M$.}
	\label{figOverallMultiple} 
	\vspace{-0.2in}
\end{figure}

\begin{figure*}[tbp]
	\centering
	\subfloat[Length.]{
		\label{length} %% label for first subfigure
		\includegraphics[width=.24\textwidth]{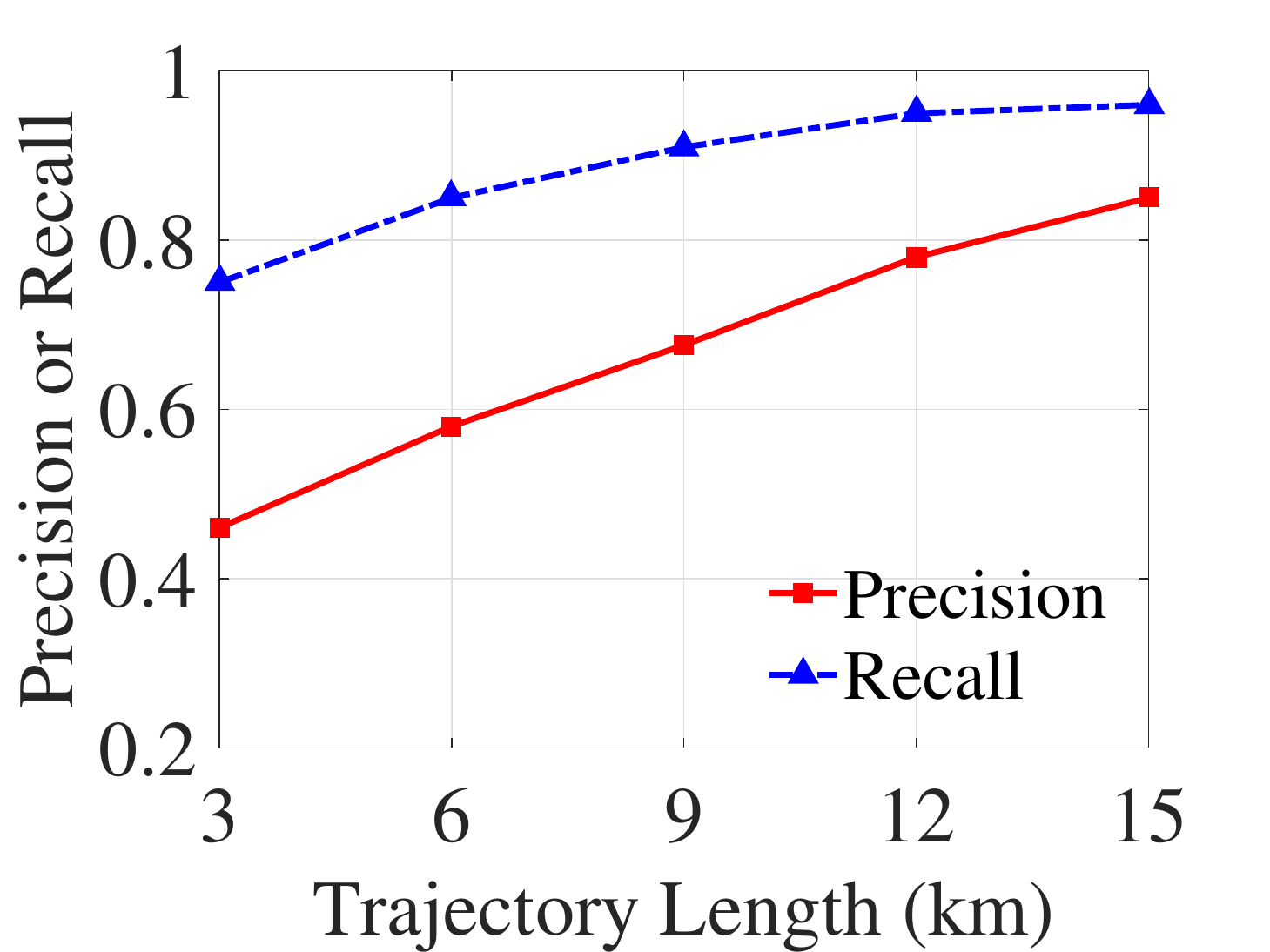}}
	\subfloat[Area.]{
		\label{location} %% label for first subfigure
		\includegraphics[width=.24\textwidth]{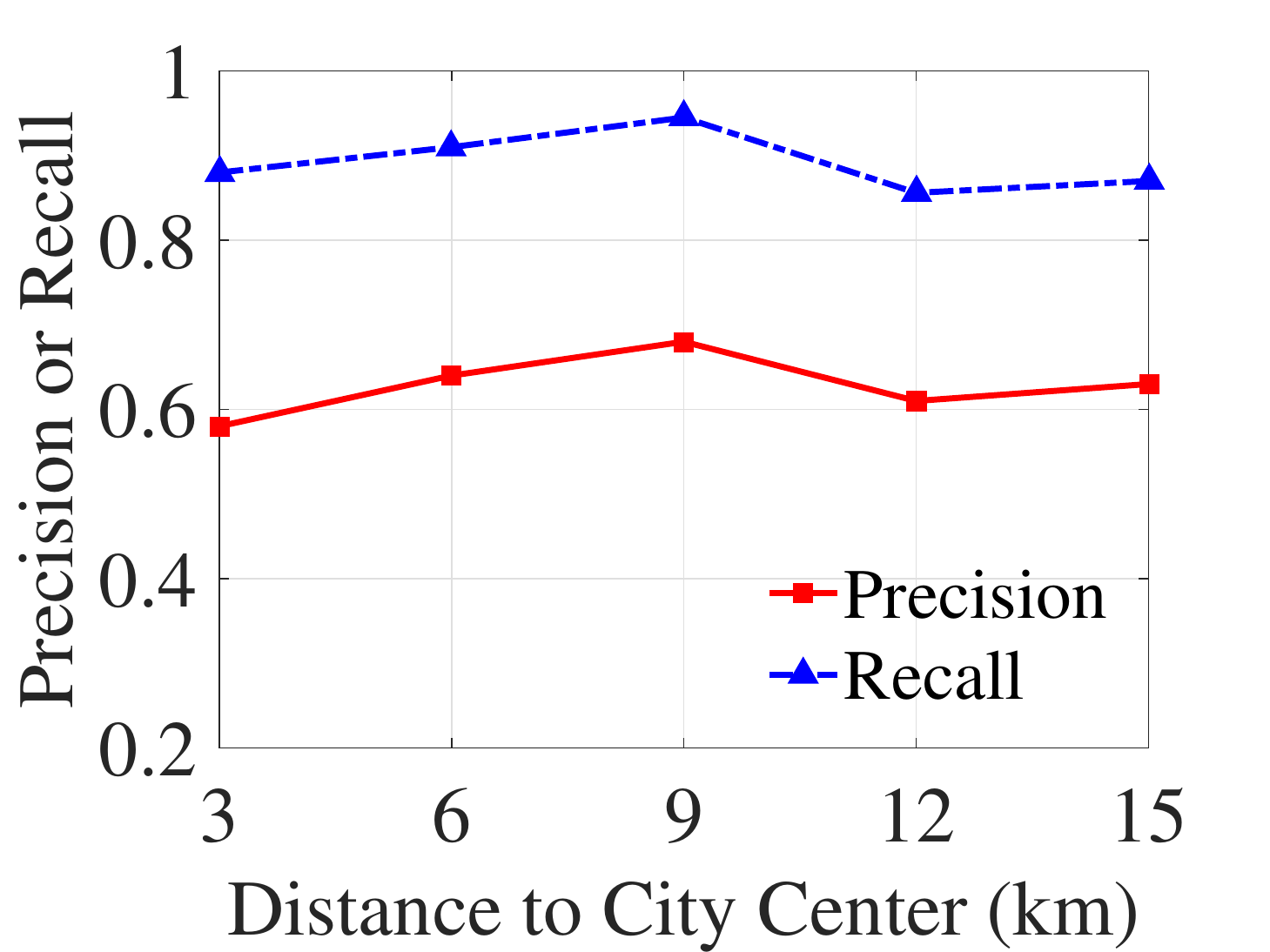}}
	\subfloat[Moving speed.]{
		\label{speed} %% label for first subfigure
		\includegraphics[width=.24\textwidth]{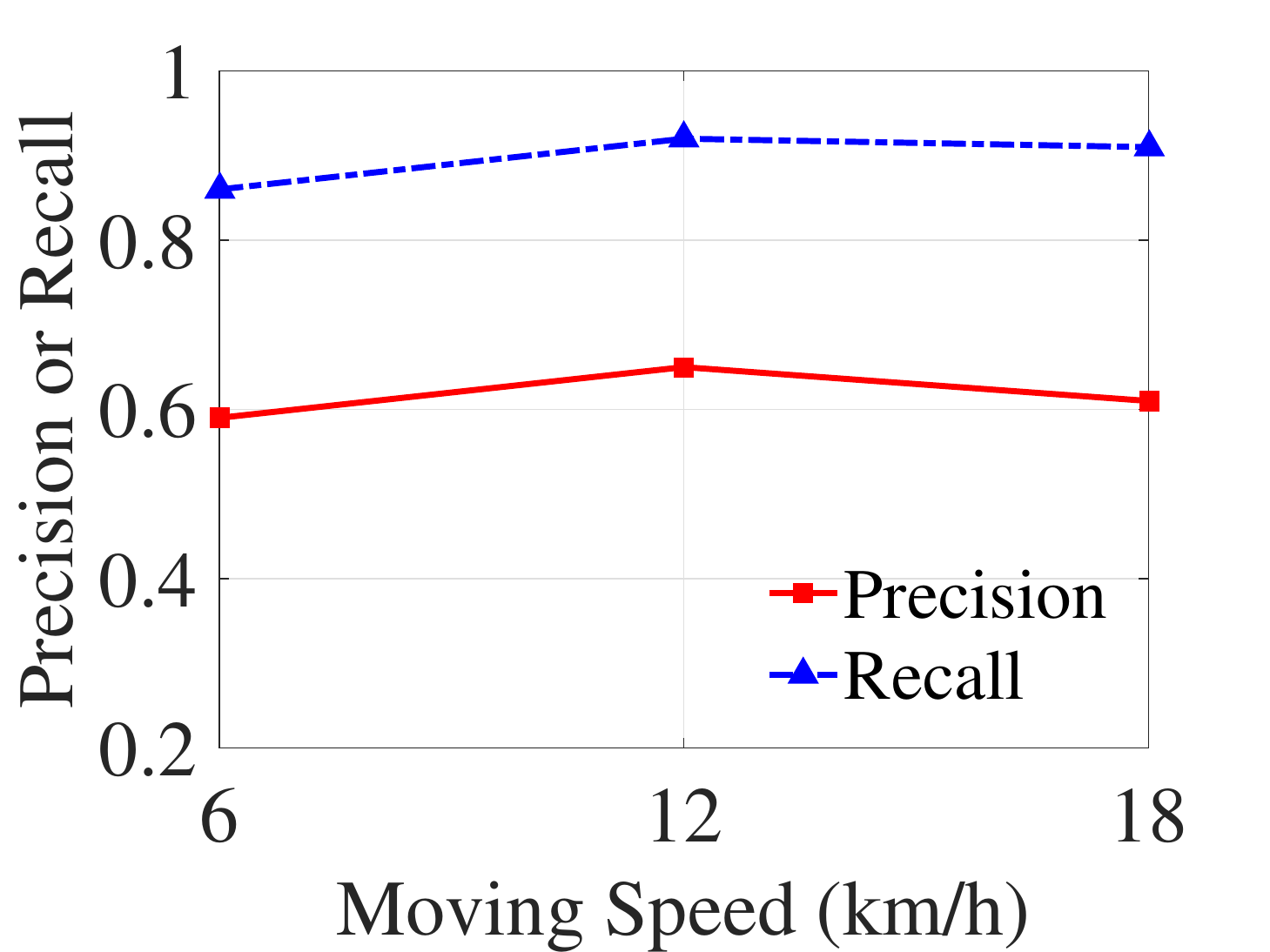}}
	\subfloat[Sample rate.]{
		\label{samplingrate} %% label for first subfigure
		\includegraphics[width=.24\textwidth]{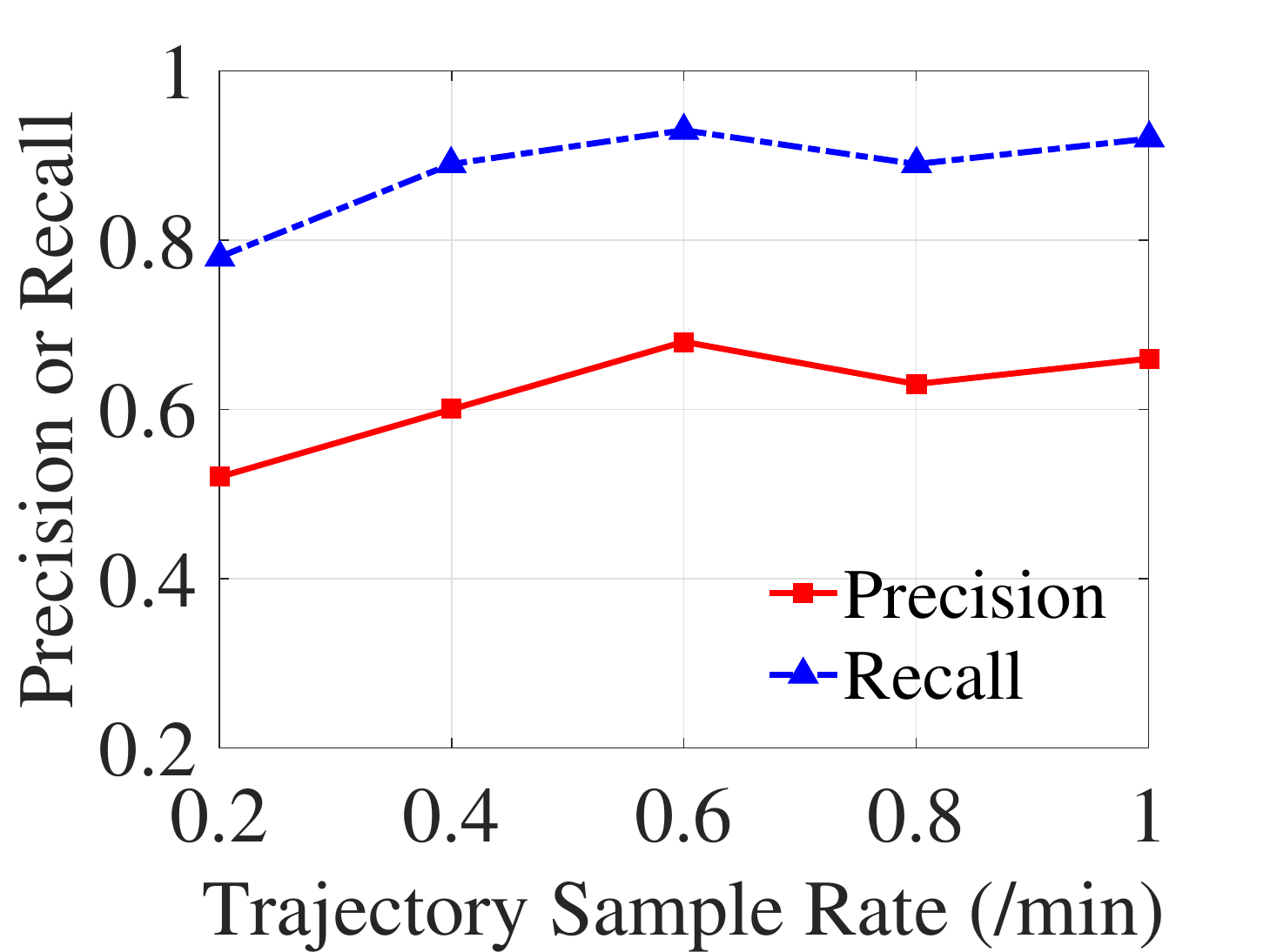}}
	\caption{Performance of cellSim under different query trajectories with different attributes.}
	\vspace{-0.1in}
	\label{figImpactOfQuery} %% label for entire figure
\end{figure*}

\vspace{-0.1in}
\subsection{Overall performance of cellSim}
\label{evaluateOverall}
We use the trajectories collected by our volunteers as ground truth to evaluate the performance of cellSim.
We first convert the cell tower sequences in our cellular dataset (including the volunteers' sequences) into trajectories using the map matching component of cellSim. Then, we use a generated trajectory or the corresponding GPS trajectory of one volunteer as the query trajectory to find the trajectories of the other volunteers in the same co-moving group from our generated trajectory dataset. 
We conduct 121 queries using different query trajectories in our collected volunteer data. Fig.~\ref{Ccomall} depicts the average experiment results of cellSim and the two baseline methods. 

We use precision and recall to study the performance of cellSim on retrieving similar trajectories. 
Precision is the ratio between the number of real similar trajectories (indicated by the ground truth) and the number of all similar trajectories identified by cellSim. 
Recall is the ratio between the number of true similar trajectories correctly identified by cellSim and the number of real similar trajectories in the ground truth.

\textbf{Cell tower sequences as query input.}
We conduct two experiments with two different cellular datasets, i.e., one with the data of multiple carriers (Fig.~\ref{Ccomall}~\subref{Cmulti}) and the other with the data of only one carrier (Fig.~\ref{Ccomall}~\subref{Csingle}).
We first transform the query input of cell tower sequences and the cellular data into trajectories by our map matching scheme. 
We then search for the similar trajectories with query trajectories from the transformed datasets. 

With the cellular data of multiple carriers, the experiment results in Fig.~\ref{Ccomall}~\subref{Csingle} reveal that cellSim can provide a high precision and recall, i.e., 62.4\% and 89.8\% respectively. CellSim can improve the precision and recall of the existing solution (SnapTS) by 88.5\% and 65.7\% respectively, based on the unique design of map matching and similarity search, and their innovative combination.
In addition, CellSim outperforms cellSim-1 by 21.6\% and 17.8\% in precision and recall respectively, benefiting from its innovative combination of map matching and similarity search.

Fig.~\ref{Ccomall}~\subref{Csingle} depicts the experiment results based on the cellular data of a single carrier. If only the cellular data of Carrier A are used,
the precision and recall of cellSim and two baselines are higher than those of multiple carriers (Fig.~\ref{Ccomall}~\subref{Cmulti}).
This is because cell tower sequences from the same carrier have lower location ambiguity.

\textbf{GPS trajectories as query input.}
The above experiments use cell tower sequences as query input. CellSim also accepts GPS trajectories as query inputs. 
Fig.~\ref{Ccomall}~\subref{Cgps} presents the performance achieved by cellSim and the two baselines using GPS trajectories as query inputs.
Compared with the experiment results with the input of cell tower sequences in Fig.~\ref{Ccomall}~\subref{Cmulti}, all three methods achieve slightly-higher performance, because the GPS trajectories on the road network have higher location accuracy in both space and time. 
% It suggest the potential applications of cellSim in the future 5G/6G mobile data, where the density of cell towers will be increased. 
On the other hand, the GPS input only mitigates the location ambiguity of query trajectories. We still need to search similar trajectories in the cellular dataset. 
As a result, in Fig.~\ref{Ccomall}~\subref{Cgps} the performance gain of cellSim over the other two baselines is still high, benefiting from its unique design for cellular data.

\vspace{-0.1in}
\subsubsection{Similarity threshold $\tau$}
Fig.~\ref{figOverallMultiple}~\subref{figTau} depicts the precision and recall of cellSim with different similarity threshold $\tau$.  
As $\tau$ decreases from 0.95 to 0.8, the recall increases, but the precision gradually decreases. When $\tau$ < 0.85, the precision declines sharply, because a smaller $\tau$ increases the number of similar trajectories in the output. 
To find a proper $\tau$, we use F-Measure~\cite{Fscore} to find a balance between precision and recall.  \begin{equation}\label{fmeasure}
F-Measure = 2 \times \frac{Precision \times Recall}{Precision + Recall}
\end{equation}

As shown in Table \ref{fscore}, we have the largest F-Measure 0.76 when $\tau$ = 0.85. Therefore, we set $\tau$ to 0.85 in our experiments.

\vspace{-0.1in}
\subsubsection{Number of trajectory candidates $M$}
The motivation of $M$ trajectory candidates is to increase the probability of including the true trajectory in our map matching output.
We conduct an experiment to verify it. 
We use our volunteers' GPS trajectories as ground truth.
We calculate the overlapping ratio between each trajectory generated by map matching and the ground truth.
% To consider one trajectory as the true trajectory, its overlapping ratio should be larger than a threshold (i.e., 0.85 in our evaluation).
The statistic results are shown in Table~\ref{multipleXM}. For $M$ trajectories of one cell tower sequence, we output the highest overlapping ratio in Table~\ref{multipleXM}. When $M$ is set to 7 or higher, there exists at least one trajectory in the $M$ trajectory candidates that has an overlapping ratio larger than 89\% with the GPS trajectory. 

\textit{Adaptation of the parameter $M$.} Fig.~\ref{figOverallMultiple}~\subref{figM} presents the performance of cellSim according to prefixed $M$ values or an adaptive $M$ value. 
As $M$ increases, the recall increases, but the precision decreases, because it is more likely that the true trajectory is included in the $M$ trajectory candidates outputted by map matching. When $M$ is larger than~7, the recall maintains stable. However, 7 trajectory candidates also make the precision decrease. Therefore, we need to adapt the $M$ value to achieve a better performance in both precision and recall.
The experiment results in Fig.~\ref{figOverallMultiple}~\subref{figM} demonstrate that our adaptation algorithm (denoted as $Adapt$ in the figure) can automatically approach the highest precision and recall simultaneously, which can only be achieved by different prefixed $M$ values separately.  

\vspace{-0.1in}
\subsubsection{Different query trajectories}
\label{len}
The above experiment results are the average results of cellSim for all the 121 query trajectories in our collected trajectory data.
We further categorize the 121 query trajectories into different categories according to their attributes, like trajectory length, moving speed and sample rate.
Fig.~\ref{figImpactOfQuery} presents the performance of cellSim according to different types of query trajectories. 

\textbf{Trajectory length.}
Fig.~\ref{figImpactOfQuery}~\subref{length} depicts the performance of cellSim with respect to different lengths of query trajectories. 
We divide all query trajectories into 5 levels of lengths. 
The precision and recall increase as the trajectory length increases, since a longer query trajectory provides map matching and the trajectory similarity search with more information to find the true trajectory and eliminate falsely-similar trajectories.

\textbf{Area.}
To evaluate the impact of the areas of the query trajectory on the performance of cellSim.  
We divide all query trajectories into 5 levels according to their distance to the center of the city. 
Fig.~\ref{figImpactOfQuery}~\subref{location} demonstrates that cellSim can almost achieve stable performance in different areas of the city, since cellSim adapts its map matching according to the density of cell towers.

\textbf{Moving speed.}
We also consider the impact of different moving speeds of the query trajectory on the performance of cellSim. We calculate the average moving speed of each query trajectory and categorize them into 3 levels according to their moving speed, i.e., \{$<6km/h$\}, \{$\geq6km/h \ \& <12km/h$\} and \{$\geq12km/h \ \& <18km/h$\}. As shown in Fig.~\ref{figImpactOfQuery}~\subref{speed}, the movement speed has very little impact on the performance of cellSim. This indicates that our algorithm is insensitive to the moving speed of query trajectories.

\begin{table}[tbp]
	\centering
	\caption{The F-Measure score according to different similarity threshold $\tau$.} 
	\vspace{-0.1in}
	\setlength{\tabcolsep}{5.5mm}{
	\begin{tabular}{ccccc}
		\hline 
		$\tau$&0.8&0.85&0.90&0.95\\
		\hline 
		F-Measure&0.37&0.76&0.73&0.69
		\\
		\hline
	\end{tabular}}
	\vspace{-0.1in}
	\label{fscore}
\end{table}

\textbf{Sample rate.}
To analyze how the sample rate of query trajectories impact the performance of cellSim, we discretize the sample rate into five levels, as shown in Fig.~\ref{figImpactOfQuery}~\subref{samplingrate}. The experiment results reveal that cellSim achieves a high performance when the sample rate reaches 0.6 per minute. Higher sample rates result in higher performance in retrieving similar trajectories, since more finer-grained information of cellular data is provided. 

\begin{figure*}[tbp]
	\begin{minipage}[tbp]{0.24\textwidth}
		\includegraphics[width=\textwidth,height=1.2in]{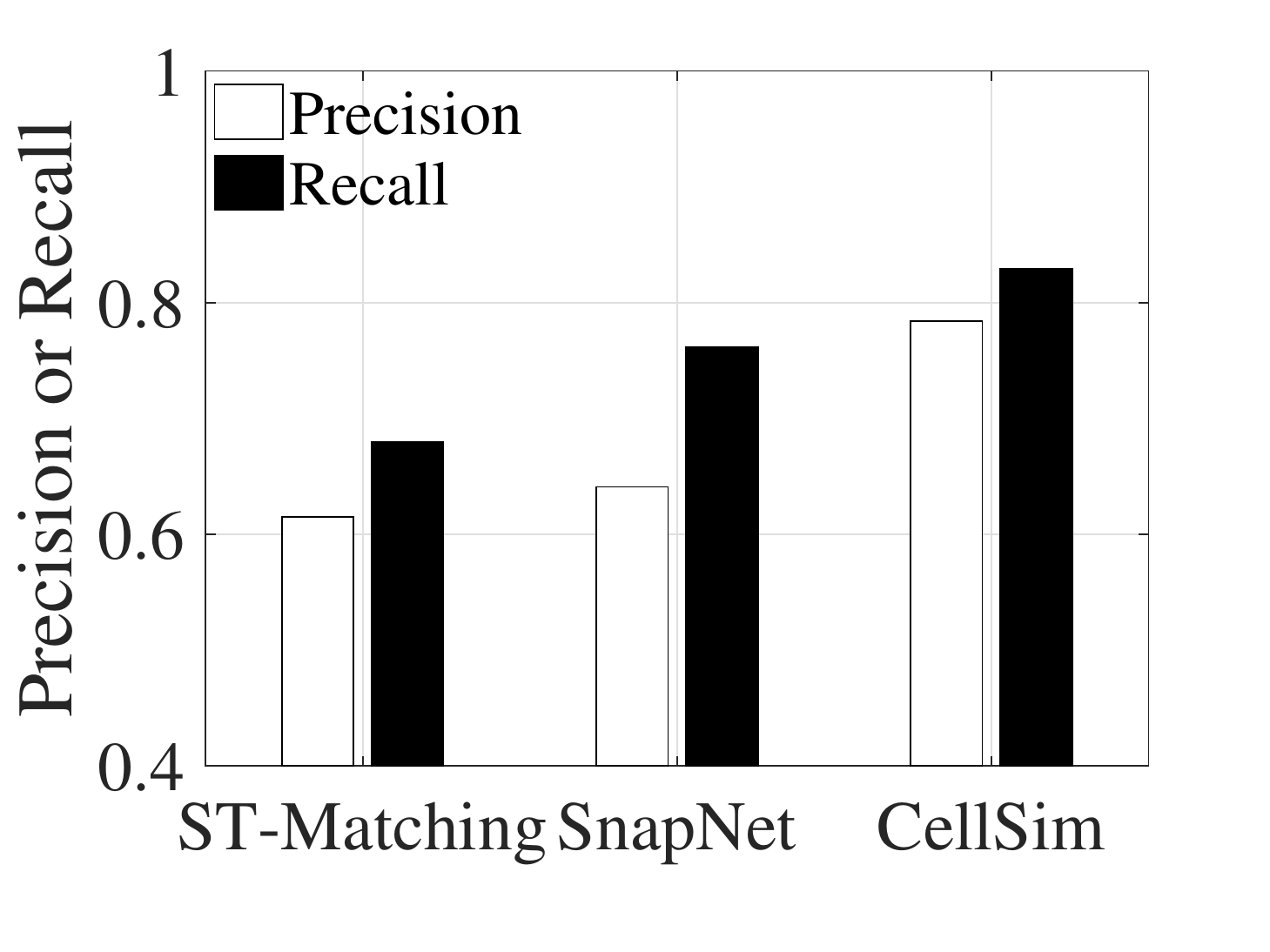}
		\vspace{-0.4in}
		\caption{CellSim map matching.}
		\vspace{-0.1in}
		\label{comparisonmapmatching}
	\end{minipage}
	\begin{minipage}[tbp]{0.24\textwidth}
		\includegraphics[width=\textwidth,height=1.2in]{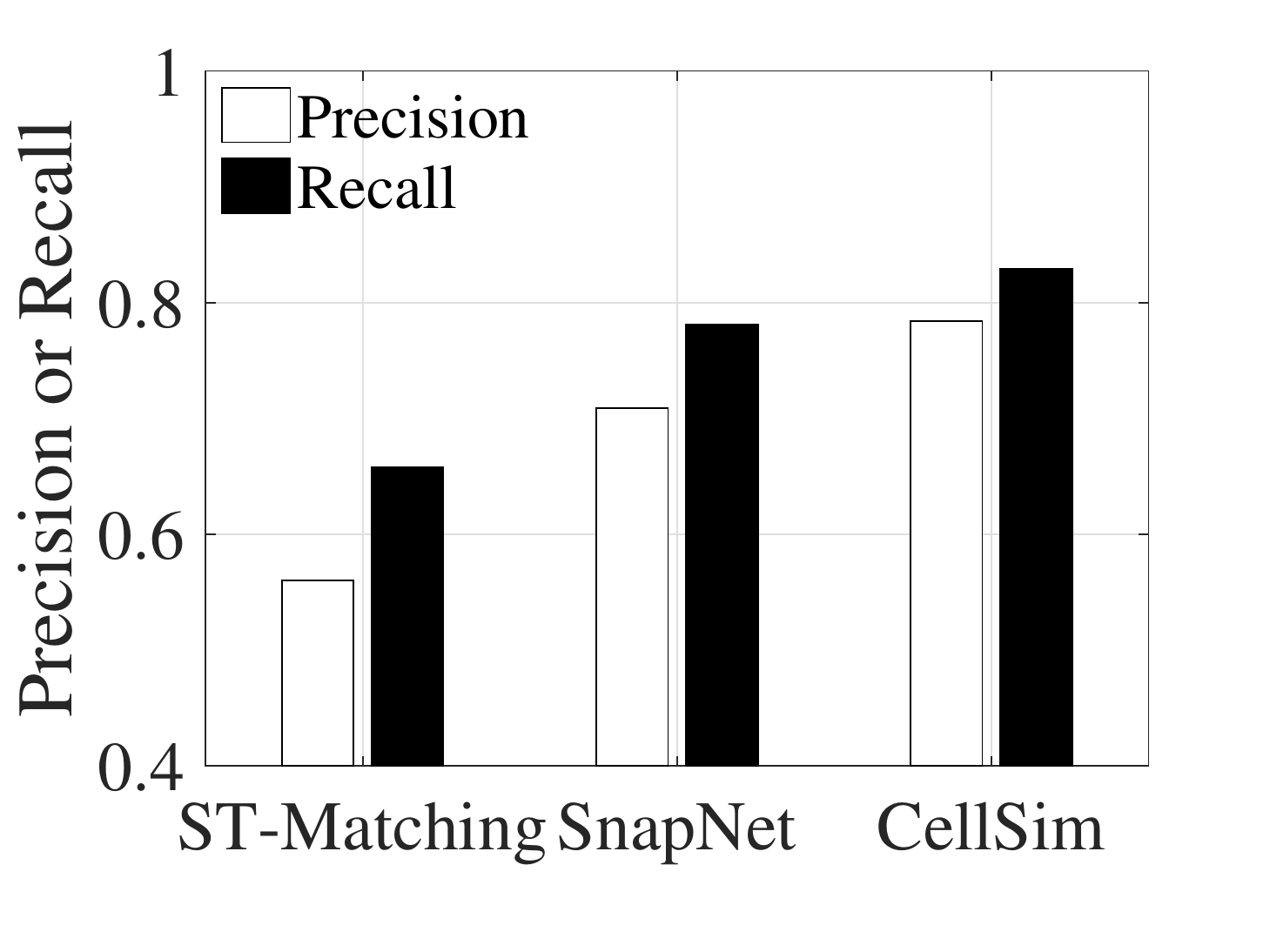}
		\vspace{-0.4in}
		\caption{Emission probability.}
		\vspace{-0.1in}
		\label{direction}
	\end{minipage}
	\begin{minipage}[tbp]{0.24\textwidth}
		\includegraphics[width=\textwidth,height=1.2in]{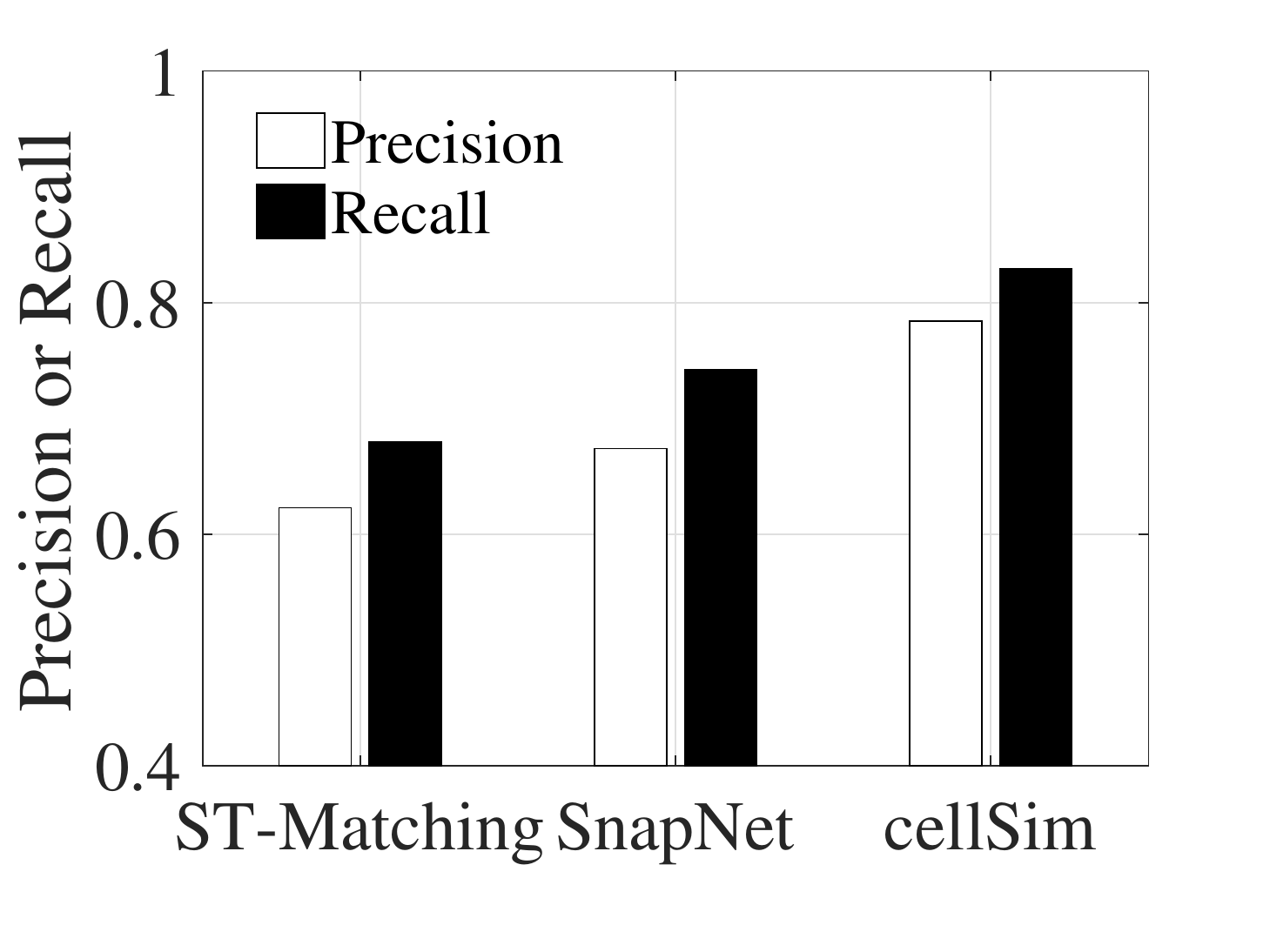}
		\vspace{-0.4in}
		\caption{Transition probability.}
		\vspace{-0.1in}
		\label{transition}
	\end{minipage}
	\begin{minipage}[tbp]{0.24\textwidth}
		\includegraphics[width=\textwidth,height=1.2in]{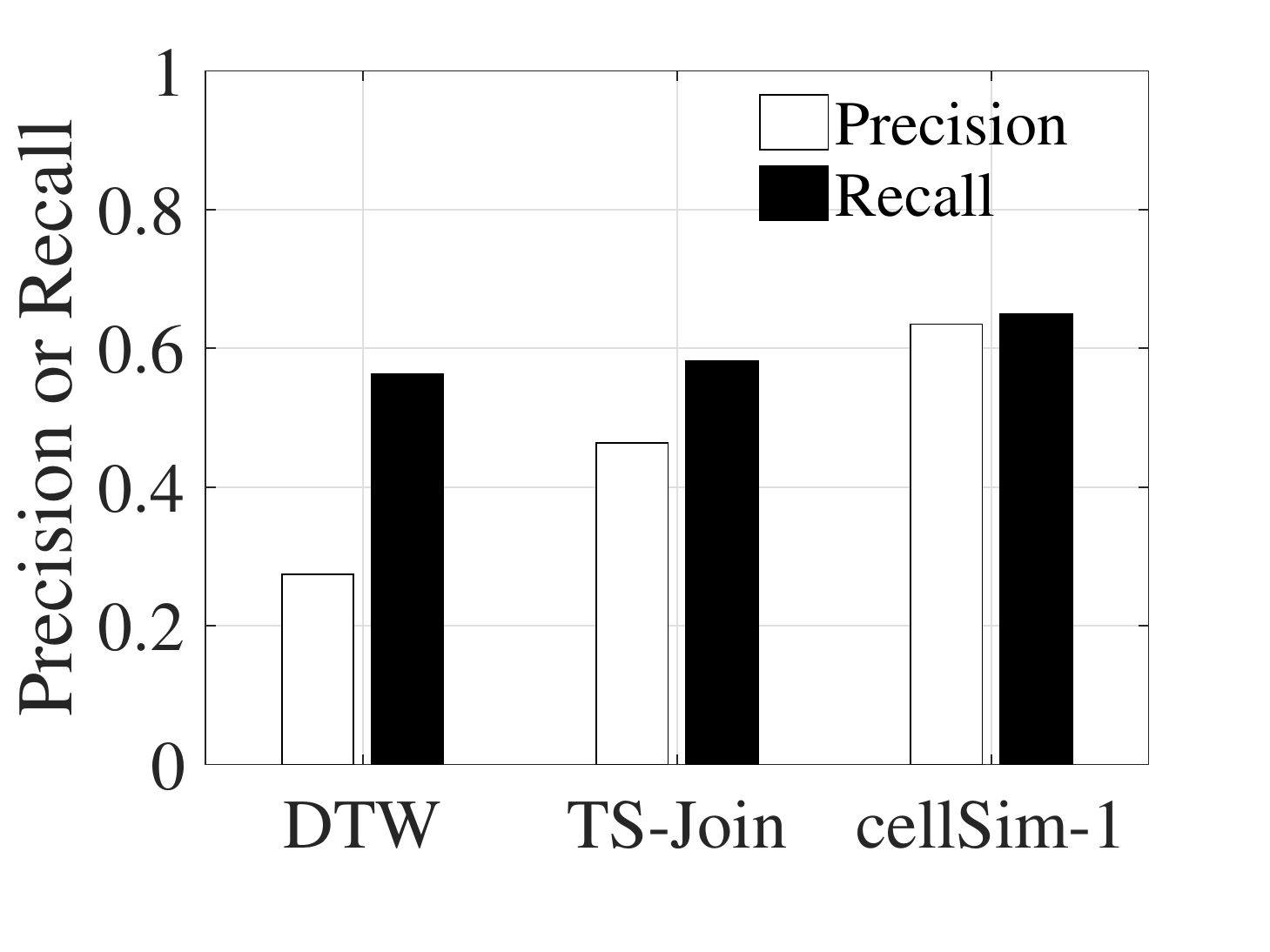}
		\vspace{-0.4in}
		\caption{CellSim similarity search.}
		\vspace{-0.1in}
		\label{Bresult}
	\end{minipage}
\end{figure*}

\begin{figure*}[tbp]
	\begin{minipage}[tbp]{0.24\textwidth}
		\includegraphics[width=\textwidth,height=1.2in]{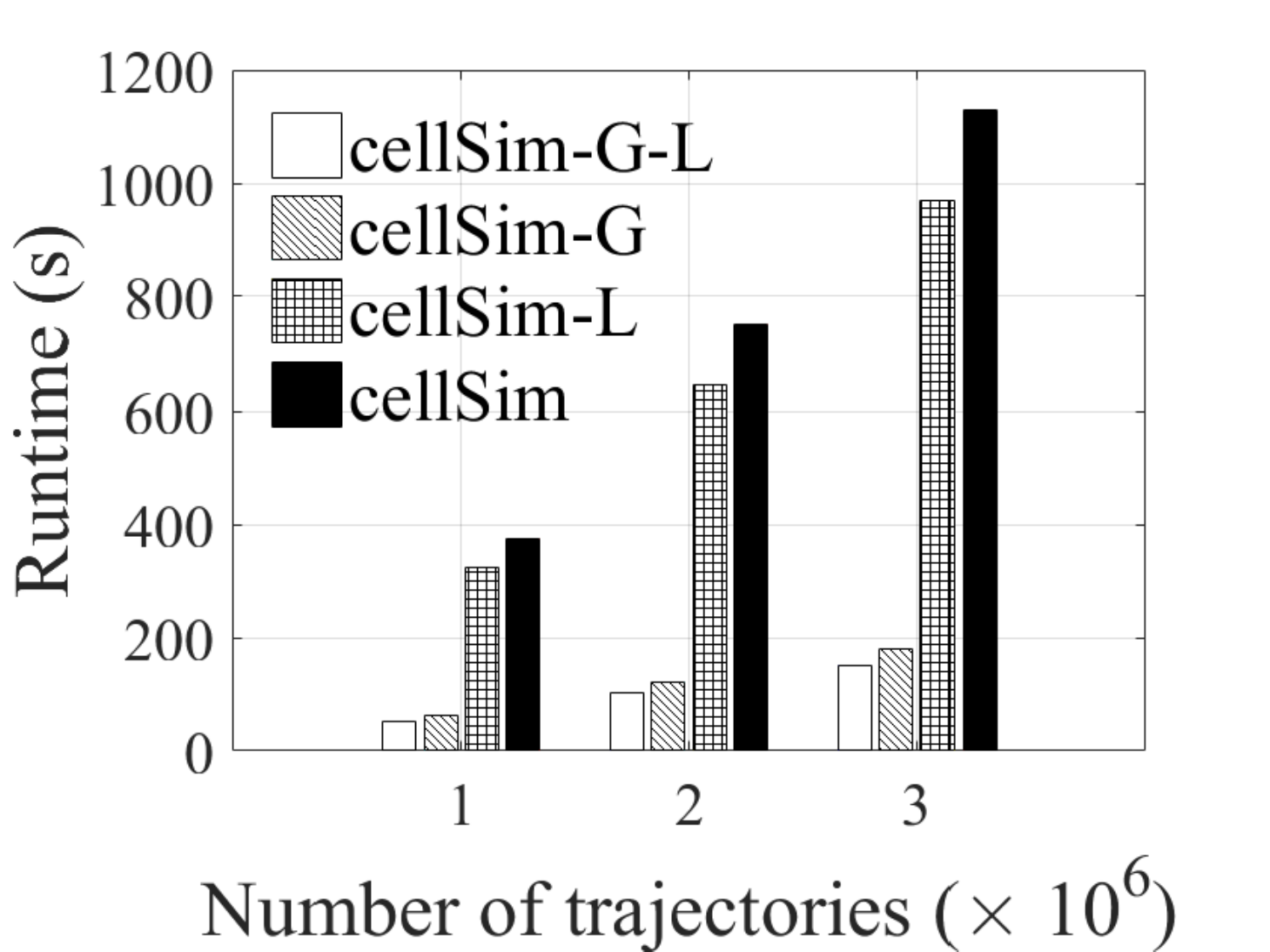}
		\vspace{-0.2in}
		\caption{Effect of pruning.}
		\vspace{-0.2in}
		\label{pruning}
	\end{minipage}
	\begin{minipage}[tbp]{0.48\textwidth}
		\includegraphics[width=0.49\textwidth,height=1.2in]{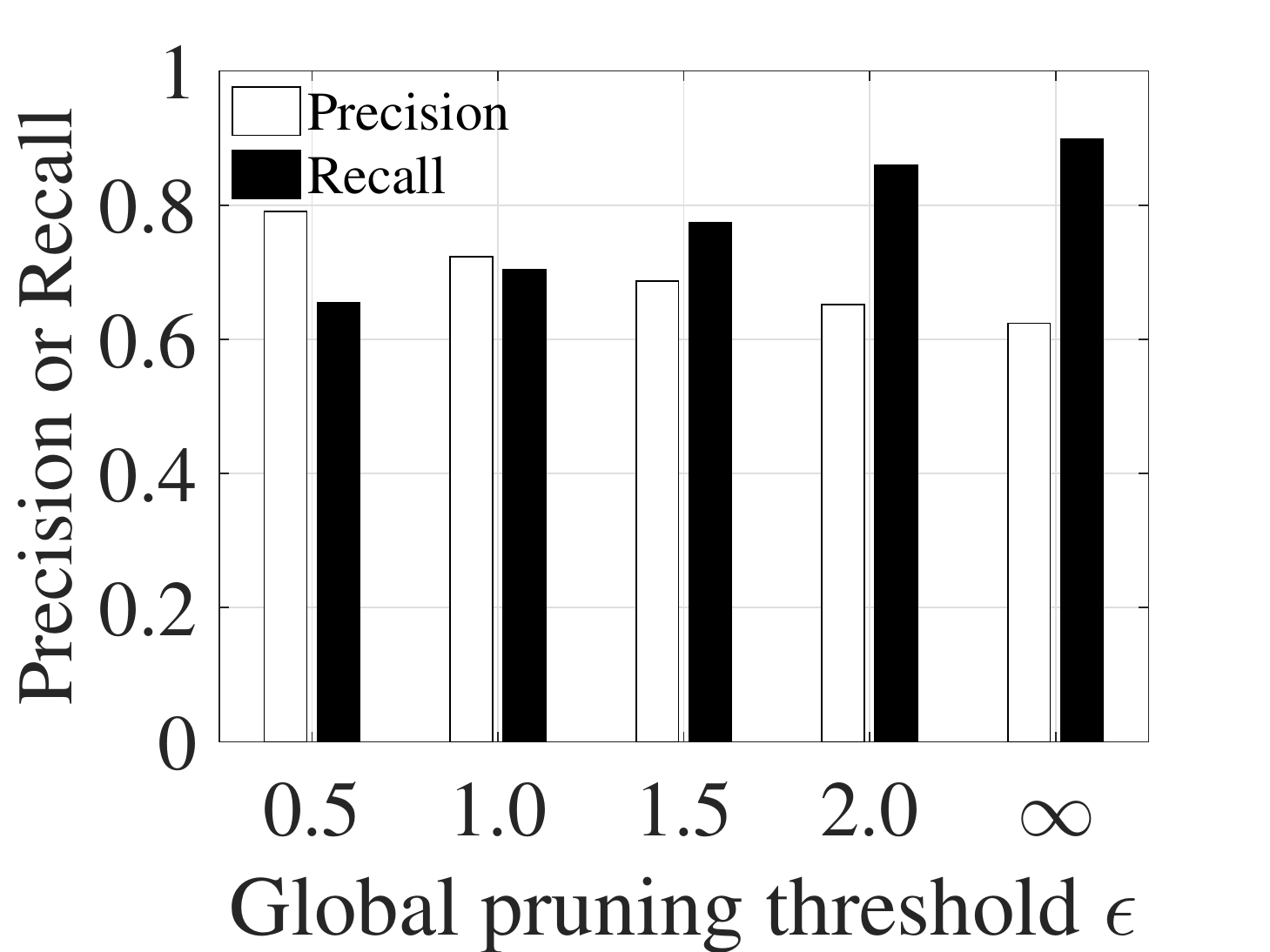}
		\includegraphics[width=0.49\textwidth,height=1.2in]{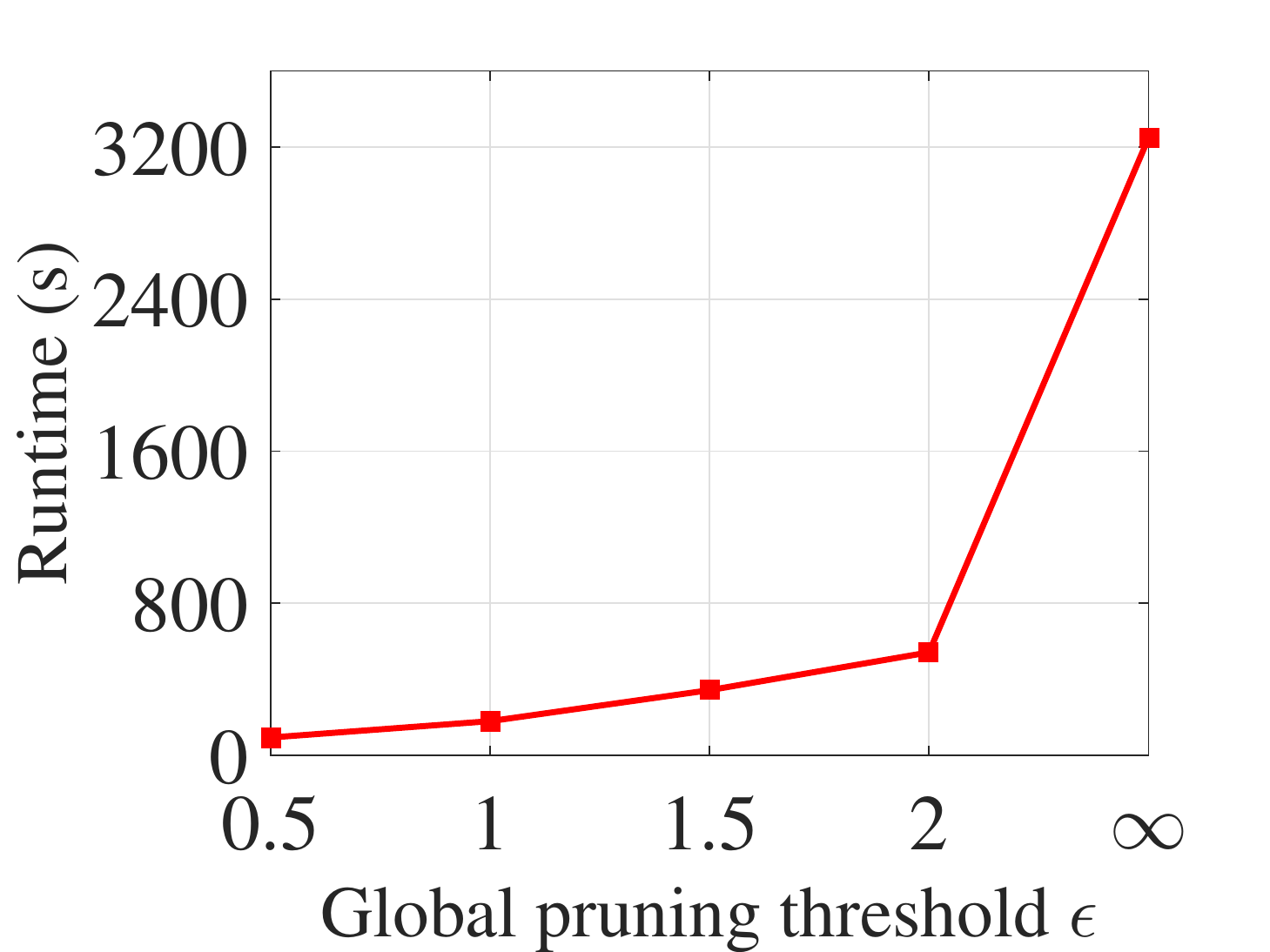}
		\caption{Effect of global pruning threshold $\epsilon$}
		\vspace{-0.2in}
		\label{global}
	\end{minipage}
	\begin{minipage}[tbp]{0.24\textwidth}
		\includegraphics[width=\textwidth,height=1.2in]{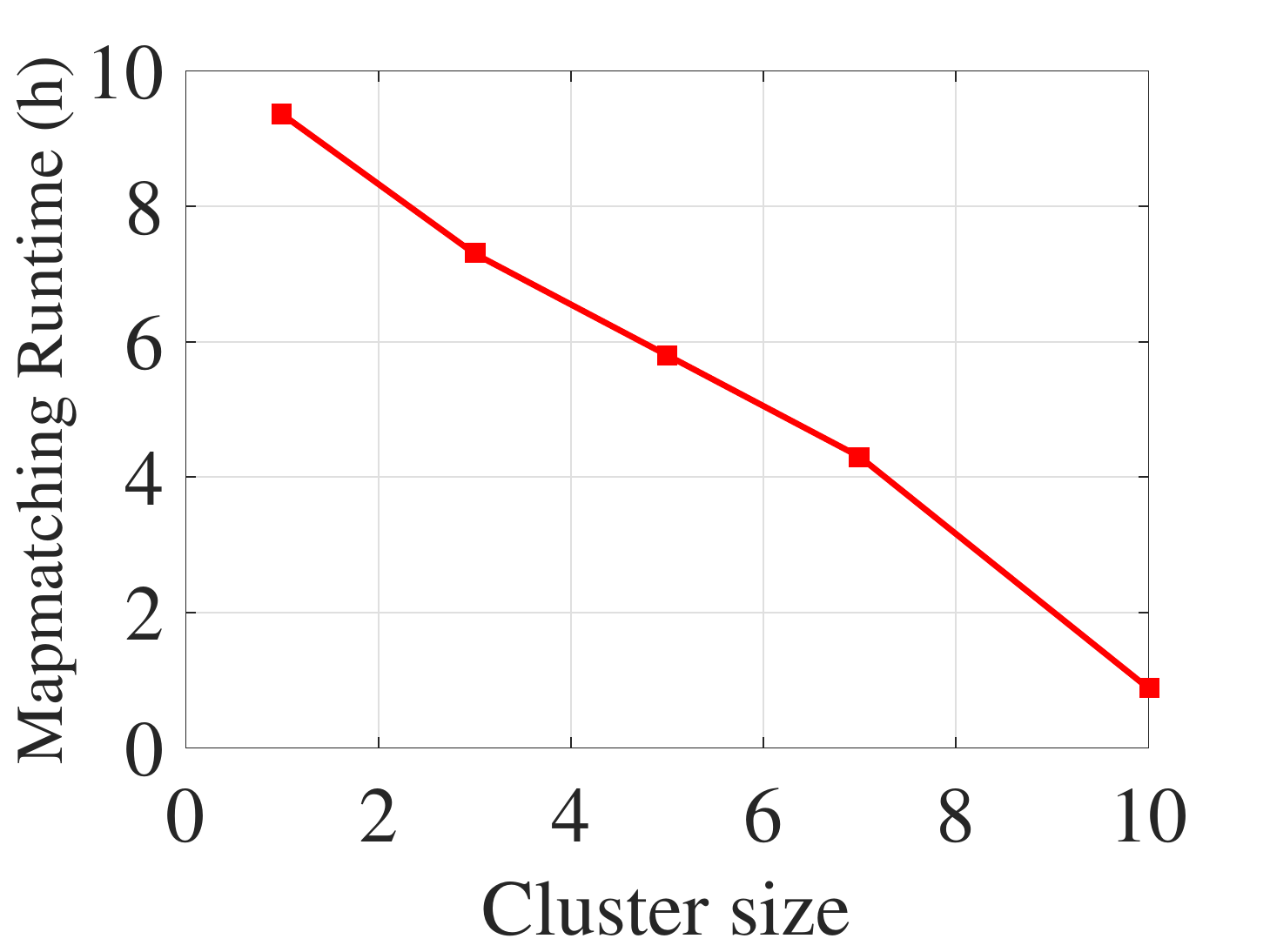}
	    \vspace{-0.2in}
		\caption{Scalability}
		\vspace{-0.2in}
		\label{mmruntime}
	\end{minipage}
\end{figure*}

\vspace{-0.1in}
\subsection{Map matching in cellSim}
We use the map matching component of cellSim to transform the cell tower sequences of our volunteers into trajectories and compare the generated trajectories with the corresponding GPS trajectories.
All the 1701-km trajectories are used in the test.
We also use precision and recall to assess the performance of map matching. 
Precision is calculated as the ratio of the total length of the correctly-matched trajectories to the total length of the whole trajectories in the map matching output. 
Recall is the ratio of the total length of the correctly-matched trajectories to the total length of the ground truth trajectories. 

As depicted in Fig.~\ref{comparisonmapmatching}, cellSim achieves the highest score in both precision and recall, 78.4\% and 82.9\%, respectively. 
The performance of SnapNet (64.1\% and 76.2\% in precision and recall) is slightly higher than ST-Matching (61.5\% and 68.0\%), because SnapNet is designed for phone-collected cellular data with higher localization errors. 
CellSim outperforms SnapNet by 14.3\% and 6.7\% in precision and recall respectively, since our map matching techniques are developed for carrier-collected cellular data with larger localization errors and lower sample rates.

% We also study the performance of using the cellular data from different carriers. 
% Figure \ref{differentcarrier} depicts that the map matching component of cellSim achieves higher precision and recall (i.e., 81.5\% and 76.4\%) using the data of Carrier A, than that using the data of Carrier B (i.e., 76.4\% and 81.3\%). 
% A possible reason is that Carrier A has more cell towers, resulting in higher localization granularity.

To further decompose the performance gain of our developed map match techniques, we test our emission probability and transition probability separately.

\begin{table}[tbp]
	\centering
	\caption{The overlapping ratio of map matching results and the ground truth.}
	\vspace{-0.1in}
		\setlength{\tabcolsep}{4.2mm}{
	\begin{tabular}{cccccc}
		\hline 
		$M$&1&3&5&7&10\\
		\hline 
		Probability &0.65&0.73&0.83&0.89&0.91\\
		\hline
	\end{tabular}}
	\label{multipleXM}
    \vspace{-0.1in}
\end{table}

\textbf{Emission probability.}
Fig.~\ref{direction} depicts the performance of different designs of emission probabilities.
CellSim can improve the map matching performance of SnapNet by 7.5\% and 4.8\% in terms of precision and recall respectively. 
In cellSim, we consider the road direction and road weight while calculating the emission probability. 

We further exploit the performance with the change of road weight $c$. As shown in Table~\ref{cweight}, we could see that our method achieves best performance at $c=$0.08. 
According to Eq.~\ref{cspeed}, by considering the road speed limit, map matching achieves the best performance at $c=$0.08.

\begin{table}[tp]
	\setlength{\tabcolsep}{1.8mm}{
		\caption{The performance with different road weight $c$.} 
		\vspace{-0.1in}
		\small
		\begin{center}
			\begin{tabular}{ccccccc}
				\hline 
				Road weight $c$&0&0.02&0.04&0.06&0.08&0.1\\
				\hline 
				Precision&0.702&0.723&0.741&0.768&0.784& 0.772 \\
				Recall&0.764&0.792&0.782&0.801&0.829&0.812 \\
				\hline
			\end{tabular}
			\label{cweight}
	\end{center}}
	\vspace{-0.2in}
\end{table}

\textbf{Transition probability.} We replace the transition probability with designs used in ST-Matching and SnapNet. It can be observed from Fig.~\ref{transition} that our transition probability achieves the highest precision and recall scores, i.e.,78.4\% and 82.9\%, compared to 64.1\% and 72.9\% by ST-Matching transition probability, and 68.3\% and 76.2\% by SnapNet transition probability. This indicates that our design for the low-sample-rate data in transition probability is effective. 

\vspace{-0.1in}
\subsection{Similarity search in cellSim}
We compare the similarity measure of cellSim with DTW~\cite{dtw} and TS-Join~\cite{vldb} under the same experiment settings (query trajectories and ground truth) used in Sec.~\ref{evaluateOverall}. We only output the top 1 trajectory candidate for each cell tower sequence using the map matching component in cellSim. In addition, we select the best similarity thresholds $\tau$ for the two baseline methods. Through many empirical experiments, we set $\tau$ to 0.85 for both DTW and TS-Join, since such a setting can achieve the highest F-measure.

As shown in Fig.~\ref{Bresult}, the similarity measure in cellSim has the best performance in both precision and recall.
First, our method adopts a time-aligned sliding window to handle the temporal dynamics of cellular data. In contrast,
TS-Join tries to find a proper parameter $\lambda$ to combine the spatial and temporal similarities, which is hard to determine in practical applications.   
Second, our method compares the spatial-overlapping ratio to capture the similarity while the other two methods use conventional distance as similarity measures. Due to the inevitable outliers in our generated trajectories caused by the large localization error of cellular data, DTW suffers from large distance differences, while our similarity measure can exclude such outliers in the overlapping process. Moreover, if we reduce the similarity threshold for larger recall, DTW introduces a high number of large false positives.

\begin{figure*}[tbp]
	\centering
	\subfloat[Ping-Pong filter]{
		\label{pingpong} %% label for first subfigure
		\includegraphics[width=.24\textwidth]{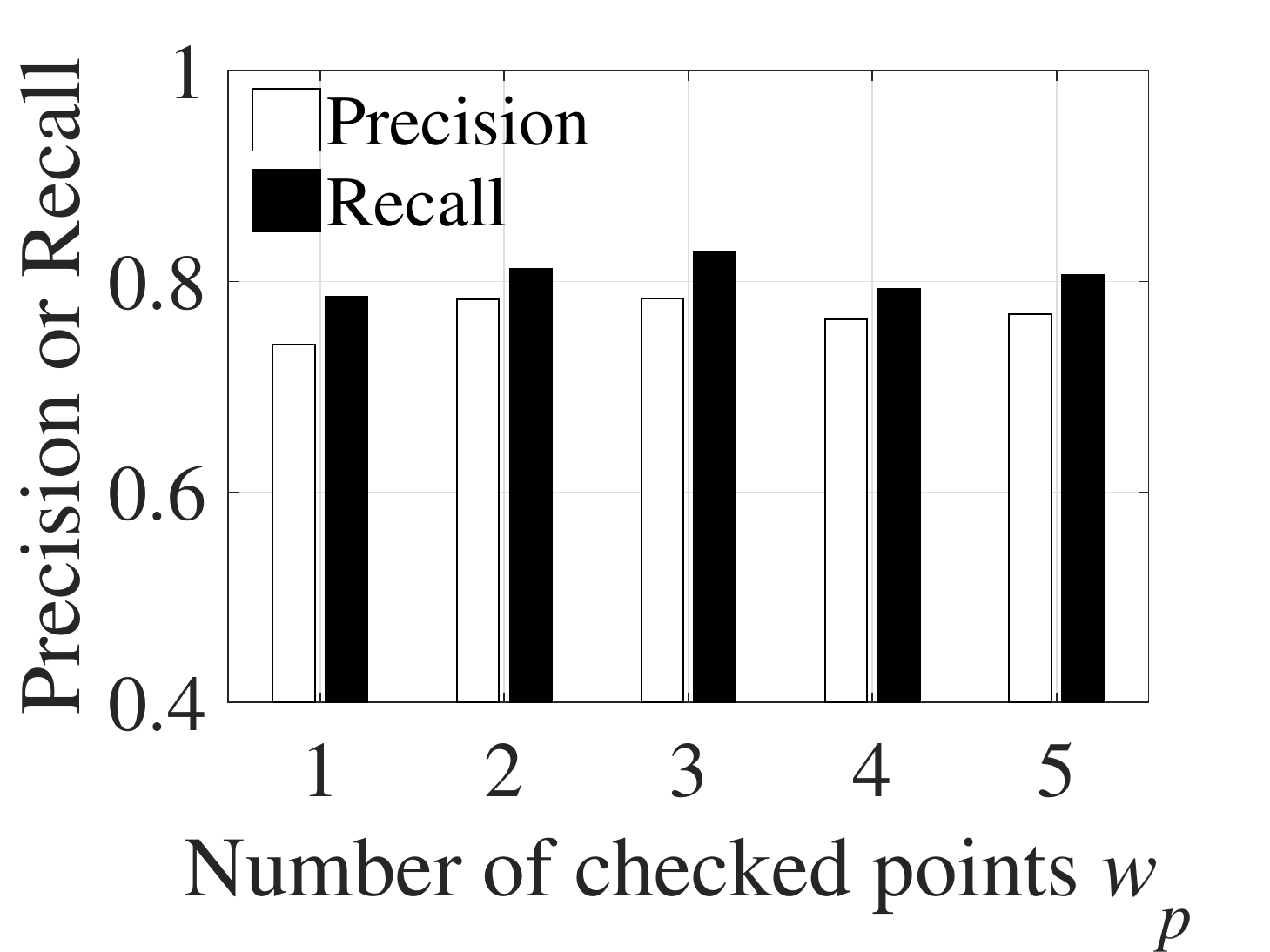}}
	\subfloat[Backward filter]{
		\label{back} %% label for first subfigure
		\includegraphics[width=.24\textwidth]{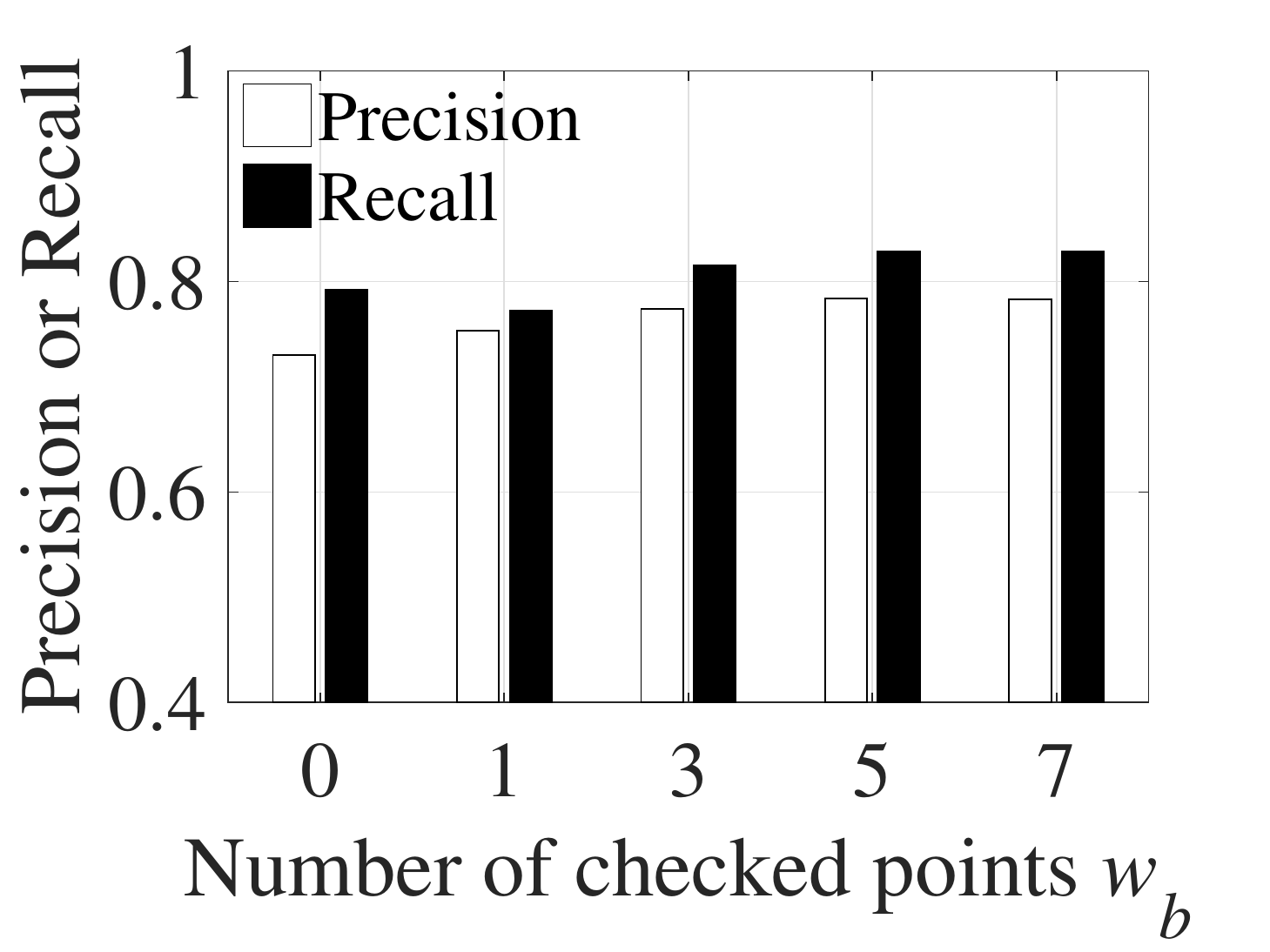}}
	\subfloat[Drifting filter]{
		\label{drift} %% label for first subfigure
		\includegraphics[width=.24\textwidth]{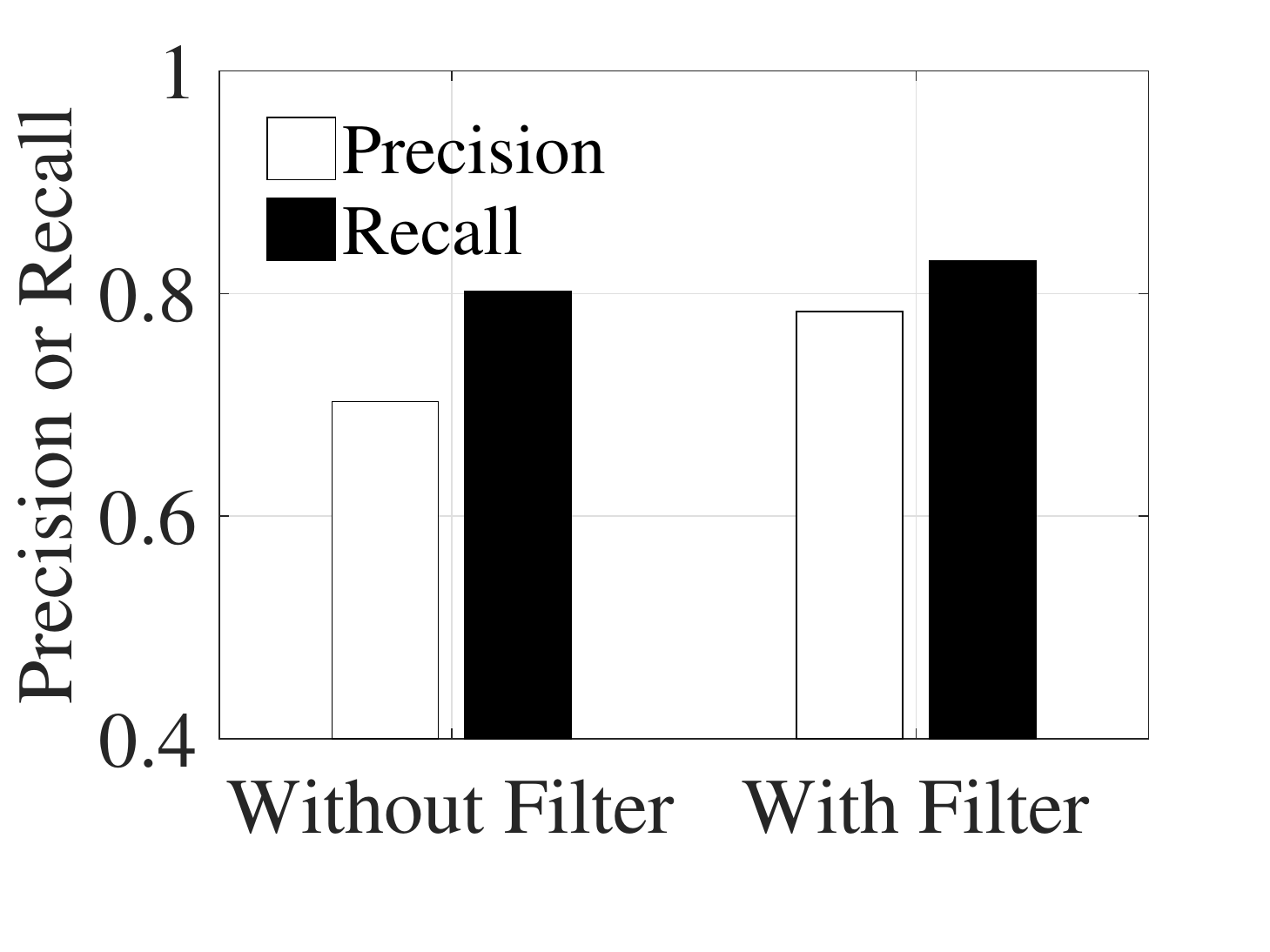}}
	\subfloat[Overall effect]{
		\label{overallpre} %% label for first subfigure
		\includegraphics[width=.24\textwidth]{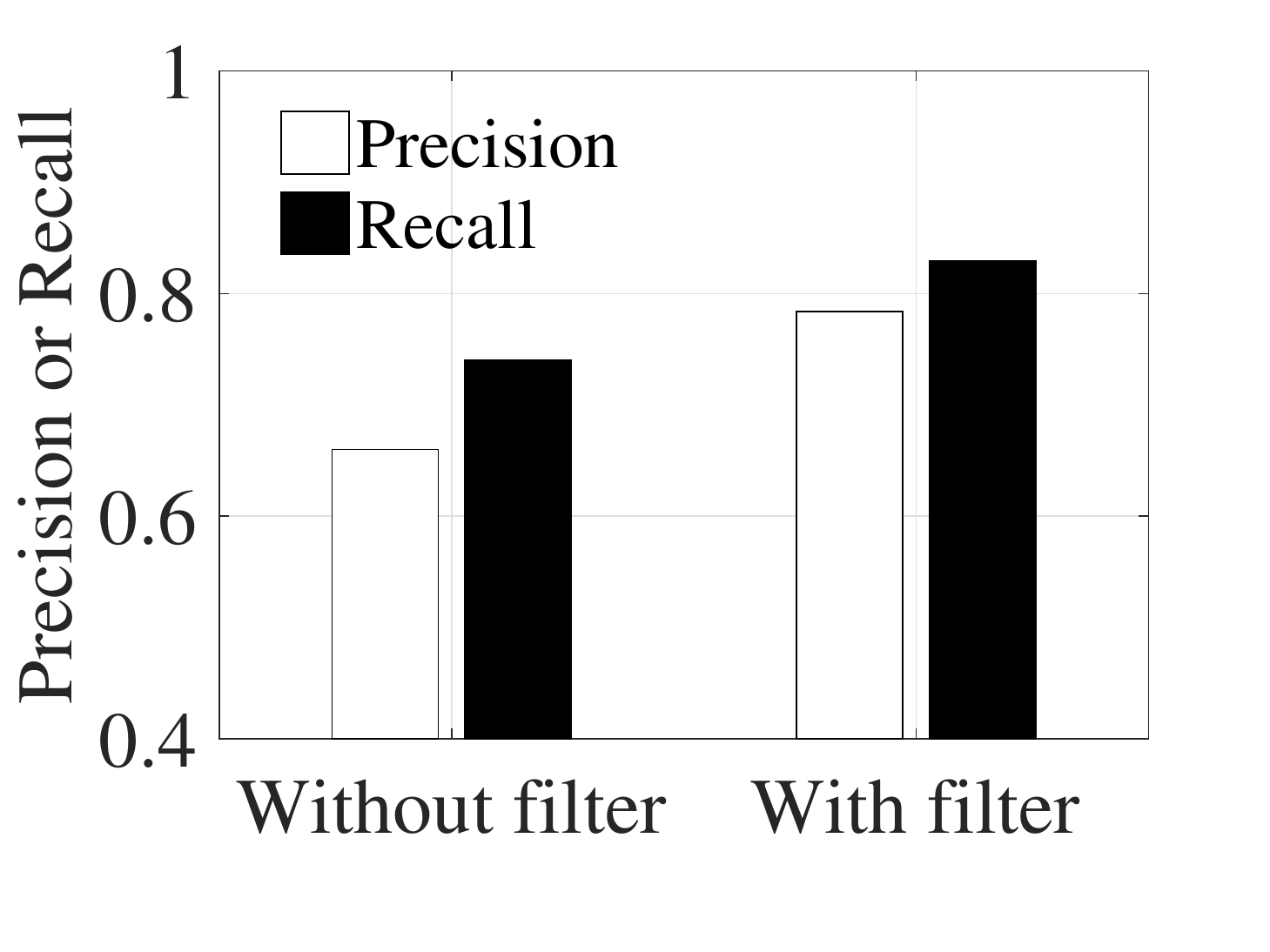}}
	\vspace{-0.1in}
	\caption{Performance of the preprocessing components in the precision and recall of map matching}
	\label{preprocess}
	\vspace{-0.2in}
\end{figure*}

\vspace{-0.1in}
\subsection{Running efficiency of cellSim}
We study the effectiveness of two pruning techniques and the scalability of the map matching component.

\vspace{-0.1in}
\subsubsection{Pruning}
We verify the  effectiveness of pruning in time saving. 
Fig.~\ref{pruning} compares the runtime by varying the number of trajectories in the dataset. 
In this experiment, we implement three pruning methods, including global pruning and local pruning, as well as both of them, denoted as cellSim-G, cellSim-L and cellSim-G-L in Fig.~\ref{pruning}. 
CellSim in Fig.~\ref{pruning} indicates that no pruning is applied in that version.
First, cellSim-G-L reduces the runtime to 152 seconds, compared to 181 seconds of cellSim-G or 969 seconds of cellSim-L when the dataset has 3 million trajectories. 
%This is because cellSim computes too many unnecessary similarity values and CellSim with two pruning methods can not only prune a great portion of trajectories in the dataset but also avoid some unnecessary comparisons in the $M^2$ comparisons. 
Second, global pruning is more effective than local pruning because it uses a global pruning threshold to avoid the computation on a large proportion of non-relative trajectories in the dataset.
Third, the performance gain provided by local pruning is higher when the number of trajectory candidates is large, because more relative trajectories need to be verified in that case and local pruning can avoid some unnecessary computation.

Fig.~\ref{global} depicts the effectiveness and efficiency of global pruning method by varying $\epsilon$ from 0.5 to $\infty$, where $\infty$ means that we do not use global pruning technique.
As we can see, when $\epsilon$ = 2, the technique can prune a large number of trajectories, resulting in less execution time, but with a slight decrease in recall. 
%We notice when $\epsilon$ becomes larger, the recall improves correspondingly, but at the same time leads to long execution time.
In addition, if we are more inclined to efficiency, we can also set a small $\epsilon$ (e.g. $\epsilon=1$). 

\vspace{-0.1in}
\subsubsection{Scalability of map matching.}
We examine how the cluster size affects the runtime of map matching. Fig.~\ref{mmruntime} depicts the runtime of map matching when the nodes of a cluster vary from 1 to 10. We discover that the process of map matching scales well and achieves nearly linear scalability. When only one node is used, 9.36 hours are taken on average to finish the map matching, compared with 1.88 hours for 10 nodes. This indicates that we can handle the incoming data in a timely manner with enough nodes.

\vspace{-0.1in}
\subsection{Data preprocessing}
We evaluate the contributions of three proposed noise filters on the map matching performance.

\textbf{Ping-Pong filter.}
Fig.~\ref{preprocess}~\subref{pingpong} depicts the performance of the Ping-Pong filter according to the number of checked points, $w_p$. When $w_p=1$, the Ping-Pong filter is not applied. 
From the experiment results, we see that the Ping-Pong filter offers a performance gain of precision and recall up to 4.4\% and 4.3\% respectively, when $w_p=3$. 

\textbf{Backward filter.}
We test the performance of the backward filter according to the number of checked point, $w_b$. 
When $w_b=0$, the backward filter is turned off. As demonstrated by Fig.~\ref{preprocess}~\subref{back}, the performance of the backward filter is stable for the different settings of $w_b$, especially when $w_b$ is set to 5 or higher.
When $w_b=1$, the accuracy is lower than the other cases, because the filter removes all backward data without the confirmation of the direction change.  

\textbf{Drifting filter.}
Fig.~\ref{preprocess}~\subref{drift} depicts the effect of the drifting
filter on accuracy. As depicted, the filter improves the precision and recall of map matching by 8.1\% and 1.3\% respectively.  

\textbf{Overall contribution of data preprocessing.}
According to the above analysis, we use $w_p=3$, $w_b=5$ as the default parameters for the Ping-Pong filter and backward filter. As depicted in Fig.~\ref{preprocess}~\subref{overallpre}, preprocessing model improve the performance of map matching in precision and recall by 12.4\% and 8.9\% respectively.

\vspace{-0.1in}
\section{Related works}
As many public cellular datasets are released by mobile carriers, many works have been conducted for a variety of applications, e.g., human mobility~\cite{zhang2014exploring,Becker2013Human,nature2009}, urban planning~\cite{Thiagarajan2009VTrack} and mobile network analysis~\cite{Li2017A}. 
A comprehensive survey on cellular data analysis can be found in~\cite{naboulsi2016large}. 
However, to the best of our knowledge, no previous works have been done to find similar trajectories using cellular data.

\textbf{Trajectory similarity search.}  
Trajectories close to a query trajectory are retrieved for various applications. 
ATSQ~\cite{ATSQ} identifies similar activity trajectories given a query trajectory for place recommendation.
SLAM~\cite{Keyframe} computes the similarity between trajectories for indoor localization.
Many trajectory similarity measure functions, e.g., Dynamic Time Warping (DTW)~\cite{dtw}, and Edit distance with Real Penalty (ERP)~\cite{erp}, have also been designed for time-series sequences. 
Recently, TS-Join defines a new similarity function which combines spatial and temporal information in a continuous manner~\cite{vldb}.
Recently, TS-Join defines a new similarity function which combines spatial and temporal information in a continuous manner~\cite{vldb}.
However, the aforementioned studies cannot be directly used for processing cellular data, because they assume the trajectories with low localization error and high sample rate.
NSIM~\cite{tuo2017nsim} is most relevant to our study, which discovers similar trajectories based on cellular data, but it cannot support to retrieve similar trajectories at city scale due to the high time complexity of their model.
%Another line of work is in efficiently designing fast algorithms for movement pattern discovery in recent years. The problem has been variously referred to as the search for flocks \cite{flock}, swarms\cite{swarm} and so on. These studies are different from ours.

\textbf{Map matching.} 
Before cellSim, two works have explored map matching using the cellular data recorded by network infrastructure. 
SnapNet~\cite{aiji} develops a HMM model for map matching and considers the road type, i.e., preferring major roads rather than side roads. 
However, it is more capable of accommodating expressways and highways. Our HMM map matching model also works for urban areas.
$Cell^\ast$~\cite{leontiadis2014cells} first identifies the accurate stationary positions by multiple sector observations and obtains the most likely path among all paths according to the moving points between stationary points. However, the sector information is not available in our cellular dataset.

Besides the above works on cellular data, many previous map matching methods are based on GPS~\cite{Newson2009Hidden,Liu2012Mining,hu2017if} and data fusion~\cite{CRF}. 
Newson and Krumm~\cite{Newson2009Hidden} propose a HMM-based map matching algorithm for GPS trajectories. 
ST-Matching~\cite{Lou2009Map} map-matches low-sampling-rate GPS trajectories with spatial and temporal information. 
Wei~\textit{et al.} makes a comprehensive comparison of map-matching methods and takes advantages of the integration in accordance with varying sample rates.~\cite{Wei2012Fast,Wei2013Map}.
Hu~\textit{et al.}~\cite{hu2017if} incorporate multi-source information to handle many ambiguous cases.
Srivatsa \textit{et al.} ~\cite{Srivatsa2013Map} study the basic assumption of HMM and identify top $k$ shortest paths to chooses the most-likely trajectory. 
However, the above works cannot be used in our work due to the large localization error and low sample rate of cellular data.

Some studies have also been conducted about the cellular data collected by mobile phones. At one location, a mobile phone receives the beacon packets from multiple cell towers. 
CTrack~\cite{thiagarajan2011accurate} and CAPS~\cite{paek2011energy} use this type of cellular data for localization. 
% Zhou \textit{et al.} \cite{zhou2014long} develop a bus tracking system based on the crowdsourced cellular data of mobile users.
Unlike these studies, our cellular data are from mobile carriers, which only provide coarse location information by one cell tower at one time point.

\vspace{-0.15in}
\section{Conclusion}
This paper presents cellSim, a practical system that can identify a group of people traveling together from a large-scale cellular dataset. By combining map matching with the trajectory similarity search, cellSim significantly improves the detection ratio of similar trajectories and achieves low false positive. 
A set of map matching techniques and a new trajectory similarity measuring method are developed for processing cellular data. 
Extensive experiments on a large dataset and real-world trajectories in a large city demonstrate the high performance gain achieved by cellSim.

% or


%\appendices


\vspace{-0.1in}
\section*{Acknowledgment}
Zhihao Shen, Xi Zhao and Jianhua Zou are supported by the National Natural Science Foundation of China Grant No. 91746111.
%This work is supported by the National Natural Science Foundation of China (Grant No. 91746111, Grant No. 71702143), Ministry of Education \& China Mobile Joint Research Fund Program (No. MCM20160302), Shaanxi provincial development and Reform Commission (No. SFG2016789)

%\section*{REFERENCES}
\vspace{-0.1in}
\bibliographystyle{IEEEtran}
\bibliography{IEEEabrv,references}

\ifCLASSOPTIONcaptionsoff
  \newpage
\fi

\vspace{-15 mm} 
\begin{IEEEbiography}[{\includegraphics[width=1in,height=1.25in,clip,keepaspectratio]{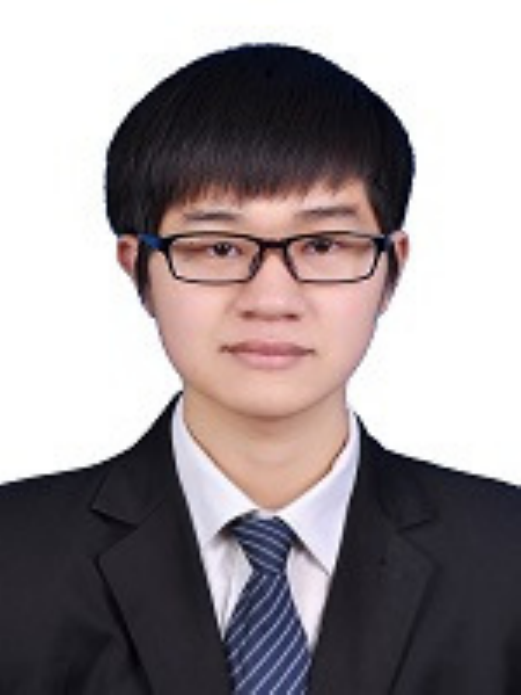}}]{Zhihao Shen} received his B.E. degree in automation engineering from School of Electronic and Information, Xi'an Jiaotong University, Xi'an, China, in 2016, where he is currently pursuing the Ph.D. degree with the Systems Engineering Institute. His research interests include big data analytics, mobile computing and deep reinforcement learning.
\end{IEEEbiography}
\vspace{-15 mm} 
\begin{IEEEbiography}[{\includegraphics[width=1in,height=1.25in,clip,keepaspectratio]{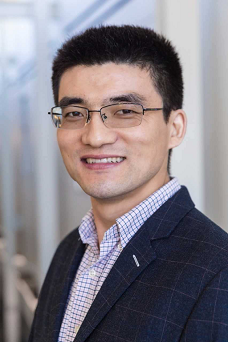}}]{Wan Du} is currently an Assistant Professor at the University of California, Merced. Before moving to Merced, Dr. Du had worked as a Research Fellow in the School of Computer Science and Engineering, Nanyang Technological University, Singapore 2012-2017. He received the B.E. and M.S. degrees in Electrical Engineering from Beihang University, China, in 2005 and 2008, respectively, and the Ph.D. degree in Electronics from the University of Lyon (cole centrale de Lyon), France, in 2011. His research interests include the Internet of Things, cyber-physical system, distributed networking systems, and mobile systems.
\end{IEEEbiography}
\vspace{-15 mm} 
\begin{IEEEbiography}[{\includegraphics[width=1in,height=1.25in,clip,keepaspectratio]{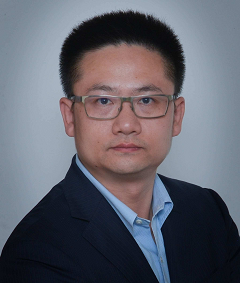}}]{Xi Zhao} was awarded his Ph.D. degree in computer science from the Ecole Centrale de Lyon, Ecully, France, in 2010. He conducted research in the fields of biometrics and pattern recognition as a Research Assistant Professor with the Department of Computer Science, University of Houston, USA. He is currently a Professor with the Xi'an Jiaotong University, Xi'an, China. His current research interests include mobile computing and behavior computing. 
% Dr. Zhao serves as a reviewer of the IEEE Transactions on Image Processing, the IEEE Transactions on Cybernetics, the International Conference on Automatic Face and Gesture Recognition, etc.
\end{IEEEbiography}
\vspace{-15 mm} 
\begin{IEEEbiography}[{\includegraphics[width=1in,height=1.25in,clip,keepaspectratio]{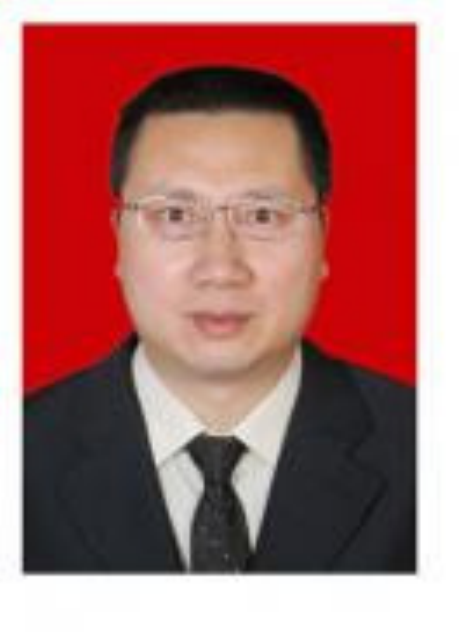}}]{Jianhua Zou}
received the Bachelor's, Master's, and Doctor's degrees from the Huazhong University of Science in 1984, 1987, and 1991, respectively. He is currently Professor in Xi'an Jiaotong University. His main research areas include: control systems and computer networks, multimedia, cognition and knowledge discovery, high voltage insulation monitoring, and complex system analysis. Since 1991, he has been the project leader and has completed about 20 research projects, including National Natural Science projects.
\end{IEEEbiography}

%\begin{IEEEbiographynophoto}{Wan Du}

%\newpage

%\begin{IEEEbiographynophoto}{Xi Zhao}

%\begin{IEEEbiographynophoto}{Jianhua Zou}


%\vfill


\end{document}